\documentclass[prd,reprint,superscriptaddress,longbibliography,eqsecnum,nofootinbib,amsmath,amssymb,aps,floatfix]{revtex4-2}

\usepackage[utf8]{inputenc}
\usepackage[T1]{fontenc}
\usepackage{multirow}
\usepackage[dvipsnames]{xcolor}
\usepackage{graphicx}
\usepackage{bm}
\usepackage[colorlinks=true,allcolors=teal]{hyperref}
\usepackage{mathrsfs}
\usepackage{leftidx}
\usepackage{orcidlink}
\usepackage{multirow}
\usepackage{siunitx}

\newcommand{\mem}{\mathrm{mem}}

\newcommand{\surr}{\mathrm{surr}}
\newcommand{\insp}{\mathrm{insp}}
\newcommand{\intr}{\mathrm{int}}
\newcommand{\rd}{\mathrm{rd}}

\newcommand{\fit}{\mathrm{fit}}

\newcommand{\model}{\mathrm{model}}

\newcommand{\UVA}{Department of Physics, University of Virginia, P.O.~Box 400714, Charlottesville, Virginia 22904-7414, USA}
\newcommand{\Obs}{Laboratoire d’étude de l’Univers et des phénomènes eXtrêmes (LUX), Observatoire de Paris, Université PSL, Sorbonne Université, CNRS, 92190 Meudon, France}

\begin{document}

\title{Waveform models for the gravitational-wave memory effect:\\
III.~Phenomenological frequency-domain model for nonspinning binaries}

\author{Arwa Elhashash\,\orcidlink{0000-0003-0379-2229}}
\email{arwa.elhashash@obspm.fr}
\affiliation{\Obs}
\affiliation{\UVA}%

\author{David A.~Nichols\,\orcidlink{0000-0002-4758-9460}}%
\email{david.nichols@virginia.edu}
\affiliation{\UVA}

\date{\today}

\begin{abstract}
We present a phenomenological frequency-domain model for the gravitational-wave (GW) memory signal from nonspinning binary-black-hole mergers on quasicircular orbits.
We develop separate amplitude and phase models for the dominant $(l,m)=(2,0)$ spherical-harmonic mode of the GW memory signal.
The amplitude and phase models are built from superpositions of elementary and transcendental functions, which can be evaluated efficiently.
Both portions of the model are calibrated against a numerical-relativity surrogate model over mass ratios from one to eight. 
Their accuracy is assessed by computing their mismatch with the numerical-relativity models using the Advanced LIGO sensitivity curve from the fourth observing run.
The mismatches are of the order $10^{-4}$--$10^{-3}$ over the parameter space of total mass covered by LIGO.
The resulting frequency-domain model is a more computationally efficient waveform model for the GW memory signal than a related time-domain model earlier produced by the authors.
An open-access implementation of both the time-domain and frequency-domain models is provided in the Python package \texttt{GWMemoryModel}, which can be used for relevant analyses of nonspinning binary black holes.
\end{abstract}

\maketitle

\tableofcontents

\section{Introduction} \label{sec:intro}

The observational field of gravitational wave (GW) astrophysics has expanded rapidly following the observations, by the LIGO-Virgo-KAGRA (LVK) collaboration, of hundreds of binary-black-hole (BBH) mergers over the past decade~\cite{LIGOScientific:2018mvr,LIGOScientific:2020ibl,LIGOScientific:2021djp,LIGOScientific:2025slb,LIGOScientific:2026wfs}.
The peak luminosity of these mergers is sufficiently high that nonlinear GW interactions produce secondary waves which have nontrivial amplitudes compared to the primary waves.
This process is responsible for producing the nonlinear (or Christodoulou) memory effect~\cite{Christodoulou:1991cr,Blanchet:1992br} (see also~\cite{Thorne:1992sdb,Frauendiener:1992dmu}).
The effect was named as such because it is associated with a change in the GW strain that remains after a burst of gravitational waves passes by a GW detector.
Both the nonlinear and the linear effect (which had been predicted earlier~\cite{Zeldovich:1974gvh}) have been shown more recently to have close connections to the asymptotic Bondi-Metzner-Sachs symmetry group~\cite{Bondi:1962px,Sachs:1962wk,Sachs:1962zza} and Weinberg's soft theorem~\cite{Weinberg:1965nx}.
See, e.g.,~\cite{Strominger:2017zoo} for a review.

Detecting the GW memory effect has several challenges.
One is the fact that the lasting offset in the GW strain is formally a zero-frequency effect, whereas GW detectors have a lower limit to their frequency sensitivity.
Therefore, GW detectors must search instead for a portion of the GW signal that is responsible for the offset associated with the GW memory, but which itself has a finite frequency content in the signal.
A common convention used in the search for the GW memory signal is the part of the full GWs that are produced by the GW nonlinearities that generate the nonlinear memory~\cite{Favata:2008yd} (though this convention goes back to calculation in the post-Newtonian (PN) approximation~\cite{Wiseman:1991ss,Kennefick:1994nw}).
Comparisons of this approximate approach to computing the GW memory to more precise NR simulations that use Cauchy-characteristic evolution~\cite{Bishop:1996gt,Bishop:1997ik,Moxon:2020gha} have shown that the approximation represents the memory signal well~\cite{Mitman:2020pbt,Mitman:2020bjf}.

A second challenge of detecting the GW memory signal is its smaller amplitude compared to that of the primary (oscillatory) GWs.
Although earlier works to detect GW signals prior to the first detection of GWs from a BBH merger~\cite{LIGOScientific:2016aoc} were more optimistic about the detection prospects from individual events~\cite{Braginsky:1987kwo,Kennefick:1994nw,Pollney:2010hs}, after the first detection, it was determined that it would be more likely that the LVK detectors would be able to detect evidence for the GW memory signal in a population of BBH mergers~\cite{Lasky:2016knh,Boersma:2020gxx,Grant:2022bla}.
Searches for the GW memory signal using LVK data have also been implemented using Bayesian evidence ratios~\cite{Hubner:2019sly,Hubner:2021amk,Cheung:2024zow} (see also~\cite{Rossello-Sastre:2026gah}) and hierarchical Bayesian inference~\cite{Cheung:2024zow,Mitman:2026zfg}.
The Bayesian approaches require evaluating a memory signal for every posterior sample of every GW event.
With the rapidly growing number of events, having efficient signal models to compute the GW memory is beneficial.

Next-generation ground-base GW detectors, such as Einstein Telescope (ET)~\cite{Punturo:2010zz} and Cosmic Explorer (CE)~\cite{Reitze:2019iox}, and space-based detectors, such as the Laser Interferometer Space Antenna (LISA)~\cite{Amaro-Seoane:2017ADS}, are more likely to detect the memory from individual BBH mergers.
See~\cite{Goncharov:2023woe} for results for ET, \cite{Grant:2022bla,Siddhant:2026vhq} for those for CE, and~\cite{Favata:2009ii,Islo:2019qht,Goncharov:2023woe,Inchauspe:2024ibs,Zosso:2026czc,Cogez:2026frh} for LISA.
Efficient and accurate waveform models are also useful for these single-event detections because of the large number of sources these detectors are expected to observe (and the corresponding stringent waveform requirements given the high signal-to-noise of these events).

In addition, there have been works showing that including the $(l,m)=(2,0)$ modes of the GW memory signal in BBH systems can improve parameter estimation by helping to resolve certain parameter degeneracies~\cite{Xu:2024ybt}.
Conversely, neglecting the full $(l,m)=(2,0)$ waveform mode can bias parameter estimation for a BBH merger with a sufficiently high signal-to-noise ratio (SNR)~\cite{Rossello-Sastre:2025gtq}.
Thus, there are an increasing number of reasons to include the $(l,m)=(2,0)$ waveform mode in GW data analysis and scenarios where it can be applied.
As a result, there are more gravitational waveform models that include the memory signal in the $(l,m)=(2,0)$ waveform mode; we review some of these models next.

When discussing the models of the $(l,m)=(2,0)$ waveform mode, it is useful to distinguish between the memory signal (which spans the inspiral, merger, and ringdown stages in the waveform) and the quasinormal-mode (QNM) contribution that arises just during the ringdown.
Several of the waveform models such as the numerical-relativity surrogate \textsc{NRHybSur3dq8\_CCE}~\cite{Yoo:2023spi}, the Phenom waveform approach~\cite{Rossello-Sastre:2024zlr,Rossello-Sastre:2025dep}, and an effective-one-body~\cite{Buonanno:2000ef} approach~\cite{Albanesi:2024fts,Grilli:2024lfh} model the complete $(l,m)=(2,0)$ waveform mode with both memory and QNM contributions to the signal.
However, as discussed in more detail in~\cite{Grant:2022bla,Elhashash:2024thm,Zosso:2026czc,Rossello-Sastre:2026gah,Siddhant:2026vhq}, having an appropriate notion of a memory signal without the corresponding contribution from the QNMs in the $(l,m)=(2,0)$ is important for correctly identifying the evidence for the memory signal in GW observations of the memory from BBH mergers.
For this reason, we previously developed a waveform model for the GW memory part of the $(l,m)=(2,0)$ mode in~\cite{Elhashash:2024thm,Elhashash:2025hqi}.

Reference~\cite{Elhashash:2024thm}, which we henceforth refer to as ``Paper I,'' took a first step towards building this model by computing the memory signal for nonspinning extreme mass-ratio inspirals and using this result to calibrate a model for the final memory offset for nonspinning binaries of any mass ratio.
The subsequent paper~\cite{Elhashash:2025hqi}, which we refer to as ``Paper II,'' incorporated the results of Paper I and constructed a time-domain waveform model for the GW memory signal using a PN solution during the early inspiral, a QNM-based solution during the ringdown, and a phenomenological waveform ansatz during the late inspiral and merger stages that interpolated between the two results.
The waveform model was calibrated for nonspinning BBH systems with mass ratios $q = m_1 / m_2$ with $1 \leq q \leq 8$ (where $m_1 \geq m_2$ is the more massive black hole (BH) in the binary).
The three parts of the time-domain model all have analytical Fourier transforms, so the time-domain model as a whole has an analytical frequency-domain representation.
However, as we discuss in more detail in Sec.~\ref{sec:review}, the finite degree of continuity in the time domain produces artifacts in the signal in the frequency domain; in addition, the analytical Fourier transform of the time-domain model involved special functions that are slow to evaluate.
In fact, it was noted in Paper~II that evaluating the fast Fourier transform (FFT) of the time-domain model was more efficient than evaluating the analytical frequency-domain representation of the time-domain model.

Most GW data analysis for the LVK is performed in the frequency domain (with the conventions described in~\cite{LIGOScientific:2026sit}), and thus, having a frequency-domain waveform model is often beneficial.
Having a frequency-domain model also avoids some of the subtleties that arise when performing signal processing on a signal such as the memory which has a nonzero offset at late times (see, e.g.,~\cite{Chen:2024ieh,Valencia:2024zhi,Elhashash:2025hqi}).
Consequently, there are are several reasons to improve the computational efficiency and spectral properties of the Fourier transform of the time-domain model of Paper~II.
This will be the main aim of this paper.
We next summarize our approach to this waveform modeling problem and how we organize the remainder of this paper.

\subsection{Summary and organization of this paper}

In this third paper in the series on modeling the GW memory signal nonspinning BBH mergers, we construct a phenomenological model for the dominant $(l, m)=(2,0)$ mode of the memory signal directly in the frequency domain.
The model contains both the amplitude and phase of the complex-valued frequency-domain GW memory signal.
The functional form of the amplitude and phase models no longer follow from the Fourier transform of a time-domain signal (as in Paper II), but make use of elementary and transcendental functions that capture the qualitative features of the signal, but which also can be evaluated more rapidly.
These amplitude and phase models are calibrated over a range of geometric frequencies and mass ratios ($1 \leq q \leq 8$) using the NR surrogate model \texttt{NRHybSur3dq8\_CCE}~\cite{Yoo:2023spi} (though we restrict to nonspinning binaries for our model).
After calibrating, the resulting model has a mismatch of order $\sim 10^{-4}$--$10^{-3}$, when using the Advanced LIGO (O4) sensitivity curve, for the nonspinning parameter space covered by our model.

This organization of our paper is summarized next.
In Sec.~\ref{sec:review}, we review a few relevant results for computing the memory signal: specifically, the multipolar expansion of the memory signal and the waveform modes used for computing the memory.
We review a few key features of the time-domain model of Paper~II (and its analytical Fourier transform) in Sec.~\ref{sec:time-domain}.
In Sec.~\ref{sec:freq-domain}, we describe the computation of the frequency-domain memory signal from the surrogate and discuss some of the relevant qualitative features of the frequency-domain memory signal.
Section~\ref{sec:model} presents the form of our phenomenological amplitude and phase models and describes their calibration.
Our results are given in Sec.~\ref{sec:results}, where we also compare the phenomenological model against the FFT of the time-domain model of Paper~II in Sec.~\ref{subsec:models_comparison} and assess the accuracy of the model through mismatch calculations in Sec.~\ref{subsec:mismatch}.
Our conclusions and discussion are in Sec.~\ref{sec:conclusions}.

\section{Computing GW memory signals} \label{sec:review}

In this section, we review the multipolar expansion of the GW strain, the corresponding multipolar expansion of the memory signal, and the waveform modes that are used to compute the memory signal.

\subsection{Multipolar expansion of the memory} \label{subsec:multipoleMemory}

To compute the $(l,m)=(2,0)$ mode of the GW memory signal for nonprecessing BBH systems, we use the same approach described in Paper~II.
The procedure starts by writing the multipolar expansion of the complex GW strain $h \equiv h_+ - i h_\times$ in terms of spin-weighted spherical harmonics $_{-2}Y_{lm}$. 
We denote the strain multipole moments by $h_{lm}$, such that
\begin{equation}\label{eq:strain_modes_Ylm}
    h \equiv h_+ - i h_\times = \sum_{l=2}^\infty \sum_{m=-l}^l h_{lm} (_{-2}Y_{lm}) \, .
\end{equation}
The multipole moments $h_{lm}$ are functions of the retarded time $u$, while the spin-weighted spherical harmonics $_{-2}Y_{lm}$ are functions of polar and azimuthal angles, often denoted by $(\iota,\phi)$, which describe the orientation of the binary with respect to the line of sight of the detectors.
As in Papers~I and~II, we compute the $(l,m)=(2,0)$ mode of the memory signal from just the nonlinear contribution, where just the oscillatory ($m\neq 0$) waveform modes are used to compute the memory signal.
We use the same expression as in Paper~II~\cite{Elhashash:2025hqi}, where the mode is given by
\begin{align} \label{eq:h20mem}
    h_{20}^{\mem}(u) = 
    &\frac{r}{\sqrt{6}} \sum_{l',l''} \sum_{m'=1}^{l'} (-1)^{m'} C_l(-2,l',m';2,l'',-m')\nonumber\\
    \times & \int_{-\infty}^{u} \mathrm du' \, \Re\Big[ \dot{h}_{l'm'}\dot{\bar{h}}_{l''m'}\Big]\, .
\end{align}
Equation~\eqref{eq:h20mem} uses the fact that for nonprecessing binaries $\bar h_{lm} = (-1)^l h_{l(-m)}$ and the properties of the spin-weighted generalization of the Gaunt coefficients to write the sum over positive values of $m'\leq l'$.
The sums over $l'$ and $l''$ also require that $l'$, $l''\geq 2$ because $h$ is a spin-2 field.

The spin-weighted generalizations of the Gaunt coefficients $C_l(s',l',m';s'',l'',m'')$ (see~\cite{Gaunt1929,NIST:DLMF}) arise from the integral of three spin-weighted spherical harmonics.
In the notation used in~\cite{Nichols:2017rqr}, the integral and coefficients are
\begin{align}
C_l(s',l',m'; & s'',l'',m'') \equiv \nonumber\\
& \int d^2\Omega \, (_{s'+s''}\bar{Y}_{lm'+m''})(_{s'}{Y}_{l'm'})(_{s''}{Y}_{l''m''}) .
\end{align}
These spin-weighted Gaunt coefficients can be written in terms of Clebsch-Gordon coefficients, 
\begin{align}
C_l(s',l',m';s'',l'',m'') = {} & (-1)^{l+l'+l''} \sqrt{\frac{(2l'+1)(2l''+1)}{4 \pi (2l+1)}} \nonumber\\
& \times \left< l',s';l'',s''|l,s'+s'' \right> \nonumber \\
& \times \left< l',m';l'',m''|l,m'+m'' \right> , 
\end{align}
and they are nonvanishing when the index $l$ is in the set $\Lambda$: 
\begin{align}
&\Lambda \equiv \nonumber \\
&\{\max(|l'-l''|,|m'+m''|,|s'+s'' |),...,l'+l''-1,l'+l''\} .
\end{align}
The conventions that we use for the Clebsch-Gordon coefficients are those implemented in \textsc{Mathematica}.

\subsection{Oscillatory waveform modes for the memory signal} \label{subsec:surrogate_memory}

The sum in Eq.~\eqref{eq:h20mem} should include all oscillatory modes $h_{l'm'}$ of the GW strain.
However, to good approximation one can use a small number of modes and still reproduce the full memory signal to a high accuracy.
As in Paper~II, we compute the memory signal from the same six modes used there: $(l,m) = (2,\pm 2)$, $(3,\pm 2)$, and $(2, \pm 1)$.

We use the \textsc{NRHybSur3dq8\_CCE}~\cite{Yoo:2023spi} surrogate model to compute these oscillatory modes.
The \textsc{NRHybSur3dq8\_CCE} model was calibrated to NR waveforms obtained using Cauchy-characteristic evolution and fixed to a particular Bondi-Metzner-Sachs frame~\cite{Mitman:2024uss}.
This waveform model has the full $(l,m)=(2,0)$ waveform mode (with both the memory signal and the QNM ringing), but we do not use this mode when computing the memory signal with Eq.~\eqref{eq:h20mem} (which restricts to $m>0$ modes).
Papers~I and~II, however, used the earlier \textsc{NRHybSur3dq8}~\cite{Varma:2018mmi} model, which was calibrated to NR waveforms obtained by extrapolation-based methods (see, e.g., \cite{Boyle:2019kee,Scheel:2025jct}).
The change in the oscillatory modes between these models is a substantial source of the differences between the model of this paper and that of Paper~II (the main reasons for the differences will be discussed in Sec.~\ref{subsec:propsFT}).
We review the general construction of the time-domain model of Paper~II next, where we will highlight a few features of the model that will produce some differences from the frequency-domain model of this paper.

\section{Time-domain model of Paper II and its Fourier transform}
\label{sec:time-domain}

We discuss the time-domain model of Paper~II in the first part of this section and the properties of its Fourier transform in the second part.
For ease of notation, we will henceforth drop the ``mem'' subscript or superscript on the $(l,m)=(2,0)$ mode of the waveform, and it will be understood that the mode $h_{20}$ or $\tilde h_{20}$ refers to just the memory component of the GW signal (in the sense of Sec.~\ref{sec:review}) and not the QNM portion that is also in the $(2,0)$ mode (as discussed in Sec.~\ref{sec:intro}).

\subsection{Time-domain model of Paper II}

The time-domain model of Paper~II was constructed to follow the different stages of the evolution of a BBH system starting during the inspiral, progressing to the merger, and ending in a ringdown (similar to other waveform modeling approaches).
Specifically, the $(l,m)=(2,0)$ mode of the time-domain memory signal was modeled as a piecewise function with three parts:
\begin{equation} \label{eq:h20td}
    h_{20}(t) = 
    \begin{cases}
        h^{\insp}_{20}(t) & \mbox{for} \ t<t_\intr, \\
        h^{\intr}_{20}(t) & \mbox{for} \ t_\intr \leq t \leq t_\rd, \\
        \Delta h_{20} -  h_{20}^{\rd}(t) & \mbox{for} \ t \geq t_\rd .
    \end{cases}
\end{equation}
The first part, $h^{\insp}_{20}(t)$, was written as a PN series in the PN parameter $x(t) = [M \Omega(t)]^{2/3}$, where $M = m_1 + m_2$ is the total mass of the binary and $\Omega(t)$ is the instantaneous orbital frequency, which evolves as a function of time due to gravitational radiation reaction.
The final part, $\Delta h_{20} -  h_{20}^{\rd}(t)$, consists of a constant $\Delta h_{20}$ for each mass ratio (the final memory offset) and $-h_{20}^\rd(t)$, which describes the approach of the memory signal to the offset $\Delta h_{20}$ during the ringdown stage of the waveform.
The part $-h_{20}^\rd(t)$ was obtained by performing a multimode QNM fit to the oscillatory waveform modes in Eq.~\eqref{eq:h20mem} for times $t > t_\rd$ (the starting time of the ringdown after the merger) and analytically computing the corresponding part of the GW memory signal.

The intermediate part, $h^{\intr}_{20}(t)$, was required because the 3.5~PN expression used in $h^{\insp}_{20}(t)$ was not sufficiently accurate during the later stages of inspiral (times of order $t_\intr \sim 10^3 M$ from merger) to represent the memory signal up until $t_\rd$ for the range of mass ratios $q \in [1,8]$ in the model.
This part of the model was phenomenological, in the sense that the form of the model (a superposition of constants and slowly growing exponential functions) did not follow from any fundamental description of the physics of the waveform, but was instead a convenient set of mathematical functions for fitting the waveforms over the parameter space of times and mass ratios.
It played a key role in determining the degree of continuity of the entire time-domain memory signal, because continuity of the strain and the first two derivatives were enforced at the times $t_\intr$ and $t_\rd$.
However, the finite degree of differentiability at $t_\intr$ and $t_\rd$ has important impacts on the behavior of the Fourier transform of the time-domain signal, which we discuss next.

\subsection{Properties of its Fourier transform} \label{subsec:propsFT}

The piecewise function in Eq.~\eqref{eq:h20td} can also be expressed as a sum of three functions multiplied by rectangular window functions to make the different piecewise components of the function have support only over the appropriate intervals of time where they are defined.
Given the linearity of the Fourier transform, the Fourier transform of this time-domain signal can be written as the sum of the Fourier transform of these windowed time-domain piecewise parts:
\begin{equation} \label{eq:h20tdFT}
    \tilde h_{20}(f) = \tilde h_{20}^\insp(f) + \tilde h_{20}^\intr(f) + \tilde h_{20}^\Delta(f) - \tilde h_{20}^\rd(f) .
\end{equation}
The labels ``insp,'' ``int,'' and ``rd'' on the Fourier transforms correspond to the rectangular-windowed versions of the corresponding time-domain signals in Eq.~\eqref{eq:h20td}.
The third term, $\tilde h_{20}^\Delta(f)$ is also a rectangular-windowed version of the constant $\Delta h_{20}$; the Fourier transform of this windowed constant is a function of frequency.
A short calculation performed in Paper~II showed that it is given by 
\begin{equation} \label{eq:h20Delta}
    \tilde h_{20}^\Delta(f) = \frac{\Delta h_{20}}{2} \left[ \delta(f) + \frac{e^{-i 2\pi f t_\rd}}{i\pi f} \right] .
\end{equation}
The Dirac delta function is related to the nonzero mean value of the step function, and the term proportional to $1/(i2\pi f)$ is related to the discontinuity in the zero-mean portion of the remaining time-domain signal.

A calculation similar to that for the Fourier transform of a step function shows that a time-domain function that has zero mean and has a discontinuity in the $n^\mathrm{th}$ derivative has a Fourier transform that scales as $1/(i2\pi f)^{n+1}$ in the frequency domain at large frequencies.
This differs from the Fourier transform of a smooth function, which falls off at high frequencies faster than any power of $1/f$.
Given that the time domain model in Eq.~\eqref{eq:h20td} has a discontinuity in the third derivative at the points $t_\intr$ and $t_\rd$, this implies that the high-frequency behavior of the Fourier transform in Eq.~\eqref{eq:h20tdFT} will scale as $1/(i\pi f)^4 = 1/(\pi f)^4$.
Thus, there will be a frequency above which the Fourier transform of the time-domain model will deviate from the frequency-domain representation of the smooth GW memory signal.
This will occur in both the amplitude of the complex $\tilde h_{20}(f)$ (where it will scale like $1/f^4$) and the phase (where it will approach zero (modulo $2\pi$).
We will highlight these features in a quantitative comparison of the Fourier transform time-domain model of Paper~II to the phenomenological frequency-domain model of this paper in Sec.~\ref{subsec:models_comparison}.

Instead, if we model the GW signal using smooth functions in the frequency domain, we will not encounter these artifacts related to the finite degree of continuity of the derivatives of the time-domain model.
To perform such modeling, we will need to understand the relevant features of the GW memory signal in the frequency domain.
This will be the next subject of this paper.

\section{Frequency-domain GW memory signal} \label{sec:freq-domain}

We first review how we evaluate the frequency-domain GW memory signal from the surrogate model before discussing the qualitative features of the amplitude and phase of the frequency-domain memory signal.

\subsection{Discrete Fourier transform of the memory signal} \label{subsec:FFT}

Our approach to computing the memory signal in the frequency domain is similar to that described in Paper~II (aside from the fact that we use the \textsc{NRHybSur3dq8\_CCE} surrogate model to compute the oscillatory ($m>0$) waveform modes that enter into the integrand for the memory).
Specifically, we use the Fourier transform integral theorem to write the Fourier transform of the memory signal in terms of its time derivative (namely the integrand in the memory integral) and the final memory offset:
\begin{equation} \label{eq:hdot-mem}
    \tilde h_{20}(f) = \frac 12 \Delta h^\mathrm{fit}_{20} \delta(f) + \frac{1}{2\pi i f} \mathcal F[\dot h_{20}] \, .
\end{equation}
A derivation of this result was given in Paper~II, and it holds for more general  time series with a nonzero mean value, not just the $(l,m)=(2,0)$ mode.
In the first, zero-frequency term, the expression $\Delta h^\mathrm{fit}_{20}$ is the fitting function (over symmetry mass ratio) for the final memory offset, which was constructed in Papers~I and~II.
The zero-frequency, delta-function term in Eq.~\eqref{eq:hdot-mem} related to the memory offset is below the low-frequency cutoffs of the LVK detectors.
We will not model it and instead focus on the remaining nonzero frequency memory signal.
The functional $\mathcal F$ is our notation for the (forward) Fourier transform, where we use the convention that a time-domain signal $h(t)$ has a Fourier transform $\tilde h(f)$ given by 
\begin{equation} \label{eq:FTdef}
    \tilde{h}(f) = \int_{-\infty}^{\infty} \mathrm{d}t \, e^{-2\pi i f t} h(t).
\end{equation}

We approximate the infinite-time Fourier transform in Eq.~\eqref{eq:FTdef} for a finite-duration memory signal by using a discrete Fourier transform using \textsc{NumPy}'s implementation of the FFT algorithm for real-valued time series.
We generate surrogate waveform modes of length of order $10^6 M$ to cover dimensionless frequencies as low as $M f \sim 10^{-6}$.
We sample the waveforms with a time step of $0.1M$ so as to accurately evaluate the numerical time derivatives in the memory signal that enter into Eq.~\eqref{eq:h20mem}.
This will limit the dimensionless frequencies covered to be in the interval $M f \in [10^{-6}, 5]$; however, as we discuss in more detail below, there will be other features in the Fourier transform of the surrogate memory that cause us to truncate the model at frequencies lower than $Mf = 5$.

\subsection{Properties of the frequency-domain memory signal} \label{subsec:FD_memory_signal_properties}

\begin{figure}
    \centering
    \includegraphics[width=\columnwidth]{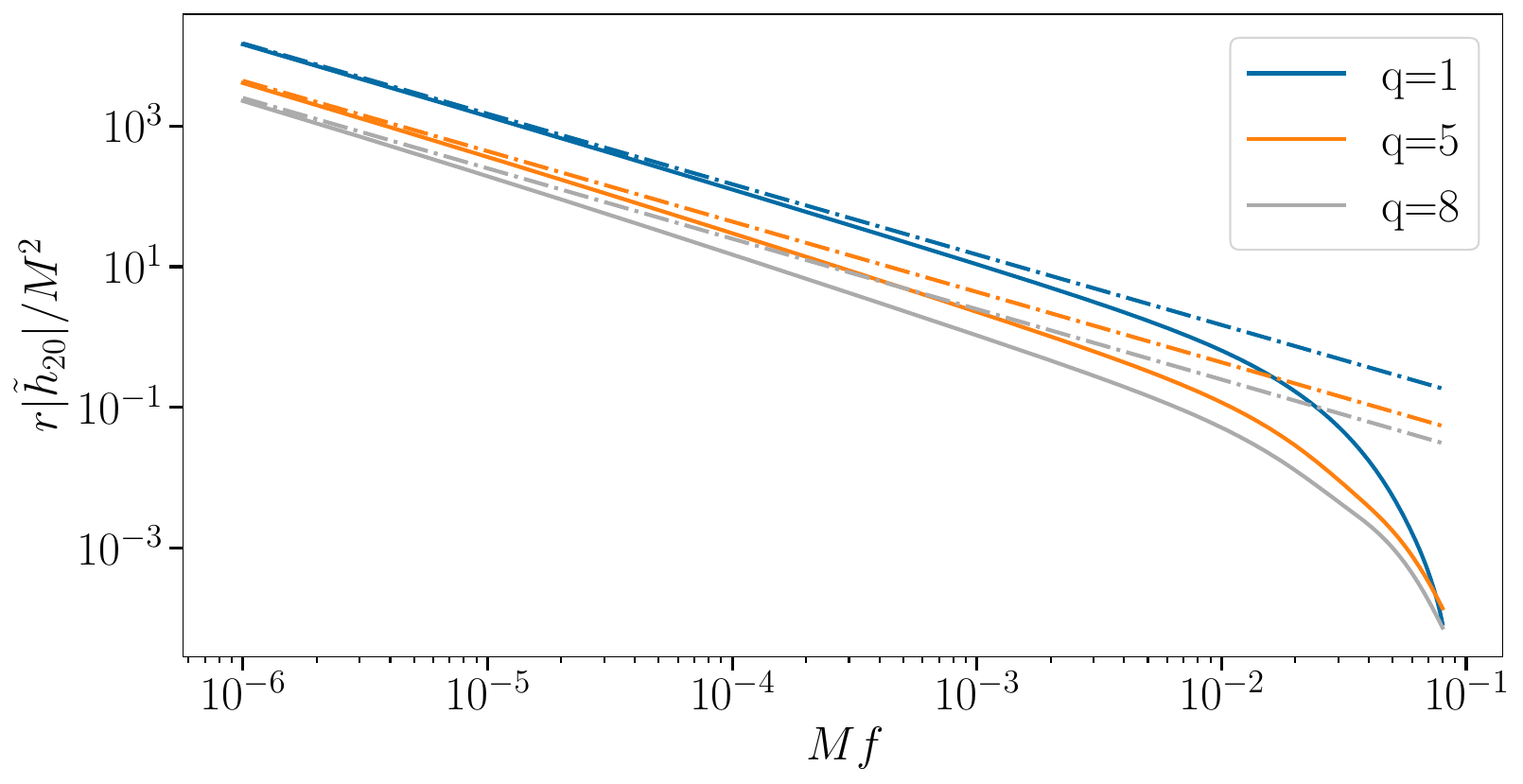}
    \includegraphics[width=\columnwidth]{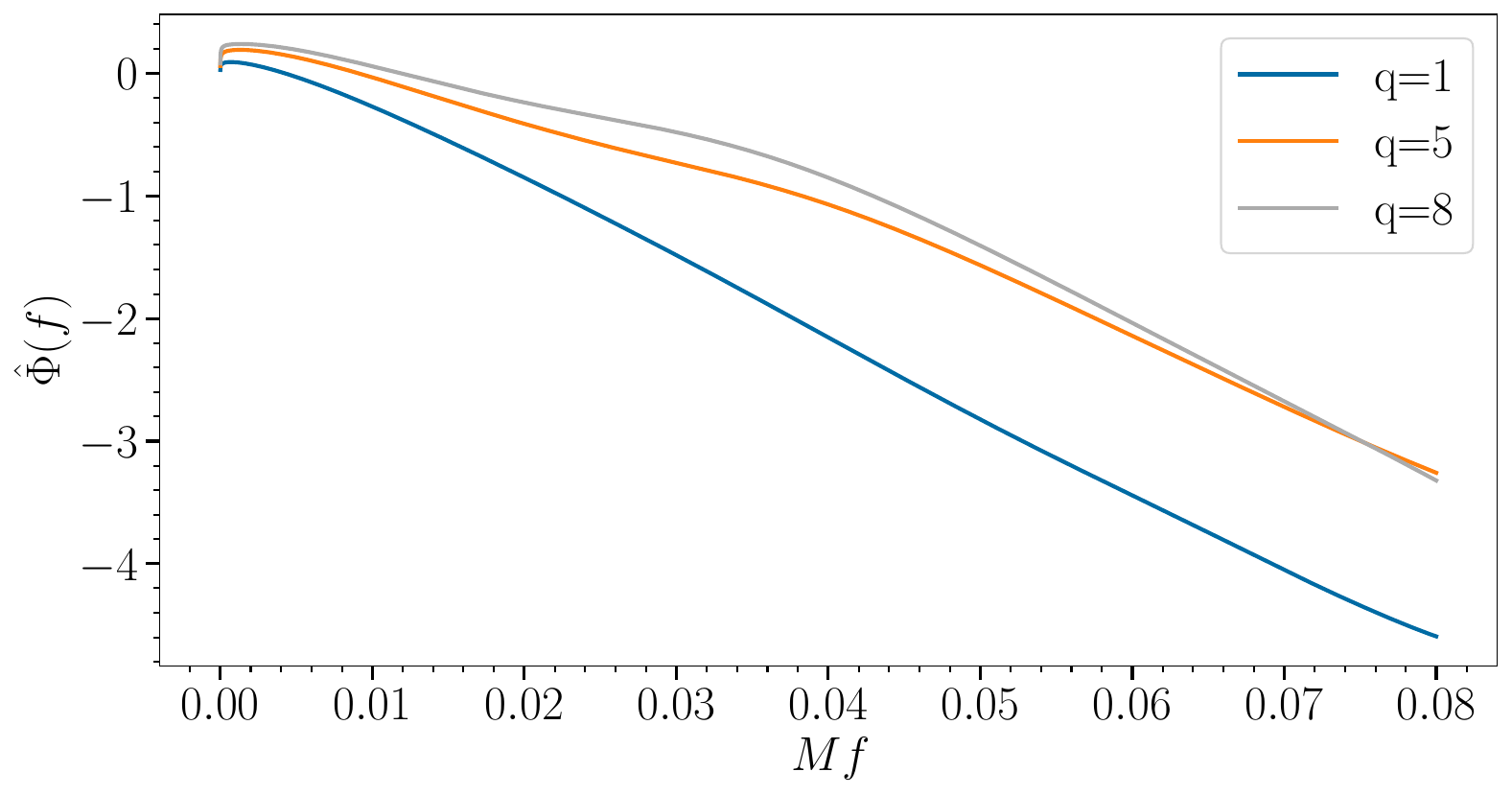}
    \caption{\textbf{Amplitude and phase of the frequency-domain GW memory signal from nonspinning BBH mergers}.
    In both panels, the blue curves correspond to a mass ratio $q=1$, the orange curves are for $q=5$, and the gray ones are for $q=8$.
    \emph{Top}: The amplitude of the frequency-domain memory signal computed from the NR surrogate model are the solid curves.
    The dashed curves are a step-function approximation to the GW memory signal in each case.
    Note that as the mass ratio increases, the convergence to the step-function approximation at low frequencies takes place at a lower frequency $M f$.
    \emph{Bottom}: The residual phase in Eq.~\eqref{eq:hatPhiMem} of the frequency-domain memory signal computed from the surrogate shown for the same mass ratios.
    The step-function approximation predicts a residual phase $\hat \Phi = 0$ (a total phase of $\Phi = \pi/2$) independent of frequency.
    The surrogate model converges to this value at low frequencies, but it deviates in a way that is described in more detail in the text.}
    \label{fig:surr_mem}
\end{figure}

Before constructing the phenomenological model, we plot the amplitude and phase of the frequency-domain memory signal and describe some of the features that our model will aim to capture.
Both the amplitude and phase of the surrogate memory signal are illustrated in Fig.~\ref{fig:surr_mem} for three mass ratios ($q=1, 5, 8$, which are displayed as blue, orange, and gray curves, respectively).
The range of frequencies in the plot is limited to $M f \leq 0.08$, because we find that there are artifacts in the amplitude and phase of the surrogate model at frequencies larger than this value, which are likely related to the finite accuracy of the surrogate interpolant.

In the top panel, the amplitude of $r \tilde h_{20}/M^2$ is shown as the solid curves and the corresponding dashed curves of the same color are the step-function approximation to the full memory signal (where $r$ is the luminosity distance).
This is computed by using the Fourier transform of a step function of amplitude equal to the final memory offset $\Delta h_{20}^{\fit}$ (from Paper~II).
Its amplitude is just $r \Delta h_{20}^{\fit}/(2\pi M^2 f)$.
The bottom panel of Fig.~\ref{fig:surr_mem} is what we define to be the ``residual'' phase, 
\begin{equation} \label{eq:hatPhiMem}
    \hat\Phi(f) \equiv \Phi(f) + \pi/2 - 2\pi t_f f .
\end{equation} 
It is related to the total phase $\Phi(f)$, which is the usual phase of a complex function: i.e.,
\begin{equation} \label{eq:h20AmpPhase}
    \tilde h_{20}(f) \equiv |\tilde h_{20}(f)| e^{i \Phi(f)} \, .
\end{equation}
The constant $t_f$ is the amount of time from the peak amplitude of $(l,m) = (2,\pm 2)$ modes of the waveform to the final time in the strain time series.
Its role will be explained in more detail below.
We discuss the attributes of the amplitude and phase of the GW memory signal and the motivation behind the definition of the residual phase $\hat \Phi(f)$ in the next two parts of this subsection.

\subsubsection{Amplitude of the strain}

Figure~\ref{fig:surr_mem} show that at low frequencies, the amplitudes of the memory signals scale as $1/f$; in addition, the signals all converge to their respective step-function approximations.
This is to be expected, because the memory offset $\Delta h_{20}^{\fit}$ was calibrated to produce the memory offset for a signal of infinite length, whereas the surrogate signal is computed from a finite-length time series.
What is perhaps less obvious to anticipate, \emph{a priori}, is that the convergence of the surrogate signal to the step-function approximation is relatively slow (as the dimensionless frequency $Mf$ approaches zero).
This convergence becomes slower as the mass ratio $q$ increases.

This slower convergence can be understood based on some of the results for extreme mass-ratio inspirals, which were discussed in Paper I.
For a fixed amount of dimensionless time ($t/M \sim 10^6$ in this case), the binary's orbital frequency increases as a function of mass ratio at the initial time.
Thus, a larger fraction of the total memory signal accumulates at times earlier than the times used in the fixed duration signal.
We chose a time-domain waveform duration of order $10^6 M$ precisely because this allowed for the step-function approximation and the full surrogate signal to agree well at the lowest frequency depicted in Fig.~\ref{fig:surr_mem}.

At higher frequencies, the amplitude deviates from the $1/f$ behavior and falls off more rapidly (what appears to exponential decay).
This is consistent with the behavior of a smooth function in the time domain (as was discussed in more detail in Sec.~\ref{subsec:propsFT}).
The precise frequency at which this turnover occurs is not straightforward to define precisely.
However, visually there appears to be a slight trend towards a deviation at lower frequencies for higher mass ratios $q$.
For example, comparing $q=8$ and $q=1$, there is a comparable relative deviation from the step-function approximation at $Mf\sim 5 \times10^{-3}$ for the $q=8$ case while for $q=1$ it occurs closer to $Mf\sim 10^{-2}$.
This behavior is consistent with the fact that binaries with higher mass ratios have a more gradual memory buildup through the late inspiral-merger, thereby lowering the characteristic frequency of the signal.
It is also related to the increased contribution from the earlier inspiral, as noted above.

For modeling the amplitude, therefore, we would like to have functions for our model that have exponential decay at high frequencies, $1/f$ behavior at very low frequencies, and a relatively slow convergence to the $1/f$ scaling in the intermediate regime.
We discuss in Sec.~\ref{sec:model} our choices for functions that have these properties.

\subsubsection{Phase of the strain}

We next discuss the phase of the surrogate memory signal.
Specifically, we show the residual phase $\hat \Phi$ in the bottom panel of Fig.~\ref{fig:surr_mem} (which uses the same color scheme for the curves as in the top panel).
First, the residual phase in Eq.~\eqref{eq:hatPhiMem} contains a term adding $\pi/2$ to $\Phi$.
This accounts for the fact that the Fourier transform of a step function has a phase of $-\pi/2$ from the factor of $1/i$ in $1/(2\pi i f)$.
Adding $\pi/2$ to the phase will make $\hat\Phi$ approach zero as the frequency approaches zero.
The second modification to $\Phi(f)$, subtracting $-2\pi t_f f$ is, in a sense, related to phase conventions that are used in evaluating the FFT (or discrete Fourier transforms, more generally).
The Fourier time-shift theorem implies that a time shift of amount $t_f > 0$ modifies the phase in the frequency domain by an additive factor of $-2\pi t_f f$.
Therefore, the amount of change in the phase of a signal as a function of frequency can be modified by a judicious time shift (or in the discrete case a circular shift).

We found that the phase of the original signal $h_{20}(t)$, which was computed with the $(l,m) = (2,\pm 2)$ modes having a peak at a time $t=0$ (namely, $t_f$ before the end of the time series), has a phase that grows as $2\pi t_f f$ as a function of frequency.
By the Fourier shift theorem, the time series which has a more slowly evolving phase as a function of frequency, $\hat \Phi$, is one for which the time series is shifted by an amount $-t_f$.
Namely, if one defines a time series $\hat h_{20}(t)$ by the inverse Fourier transform of $|\tilde h_{20}(f)| e^{i\hat \Phi(f)}$, then one obtains the relationship that that $h_{20}(t) = -i \hat h_{20}(t+t_f)$.
Thus, the $\hat h_{20}(t)$ corresponds to a purely imaginary version of the original time series that is time shifted such that the peak of the $(l,m) = (2,\pm 2)$ modes arise at the end of the time series ($t=t_f$), not at $t=0$ (along with the corresponding shift in the memory signal).
This accounts for the factor of $-2\pi t_f f$ in the definition of $\hat \Phi$.

We will model the residual phase $\hat \Phi$, and we describe its qualitative properties as a function of mass ratio.
The residual phases for a few mass ratios are shown in the bottom panel of Fig~\ref{fig:surr_mem}.
At low frequencies ($Mf \lesssim 10^{-3}$), the residual phase $\hat\Phi$ approaches zero for all mass ratios (which is consistent with the full phase $\Phi$ approaching the step-function-approximation value of $-\pi/2$).
In all cases, $\hat \Phi$ increases slightly between $Mf = 0$ and small positive values of $Mf$, before it starts to turn over as frequency increases.
At the intermediate frequencies ($1 \times 10^{-2} \lesssim Mf \lesssim 4 \times 10^{-2}$) in Fig.~\ref{fig:surr_mem}, the phase slightly decreases roughly linearly with frequency, and the magnitude of the slope is a decreasing function of the mass ratio.
In the rest of the range plotted, there is a transition to a steeper slope at frequencies $Mf \gtrsim 4 \times 10^{-2}$ for the unequal mass ratios; however, the equal mass-ratio case does not have a pronounced change in slope.

For the phase model, we will aim to construct a model that contains two different linear slopes as a function of frequency as well as the small region of increasing phase at the lowest frequencies depicted.
We will describe a model that can do so next.

\section{Frequency-domain GW memory model and its calibration} \label{sec:model}

We first discuss the mathematical form of the amplitude and phase models, and we subsequently summarize how we calibrate the memory model.

\subsection{Phenomenological amplitude model} \label{subsec:Phenom_model_amp}

The motivation for our choice of functions for the amplitude model is related to the fact that, in the time domain, the memory qualitatively resembles $1+\tanh[(t-t_0)/\tau]$, where $t_0$ is the merger time and $\tau$ is a characteristic timescale over which most of the offset in the memory signal accumulates.
The Fourier transform of a hyperbolic tangent function is known analytically and is a hyperbolic cosecant function, $\mathrm{csch}(\pi\tau f)$, plus a delta-function term at zero frequency. 
Such a function captures two aspects of the frequency-domain memory signal.
First, at frequencies $f \ll 1/\tau$, the hyperbolic cosecant scales as $ \sim 1/(\pi \tau f)$, which reproduces the $1/f$ behavior illustrated in Fig.~\ref{fig:surr_mem}.
Second, at frequencies $f \gg 1/\tau$, the hyperbolic cosecant decays exponentially as $e^{-\pi f \tau}$, which is consistent with the fact that it is a smooth function.

However, a single hyperbolic cosecant will not be sufficient to model some of the other aspects of the frequency-domain GW memory signal that we described in Sec.~\ref{subsec:FD_memory_signal_properties}.
Specifically, the hyperbolic cosecant admits a Laurent series expansion for $f \tau \ll 1$, and it would not be able to represent the slow convergence of the memory signal to the $1/f$ behavior at low frequencies (which would require fractional negative powers of $f \tau$ instead).
However, a sum of a hyperbolic cosecant and a function of the form $(\tau f)^p \mathrm{csch}(\pi \tau f)$ could capture this behavior for an appropriate power $p$.
Though such an ansatz could fit an individual mass ratio, fitting many mass ratios required introducing a third hyperbolic cosecant (specifically, so that the varying high and low-frequency behaviors of the memory signal's amplitude could be captured across the parameter space of mass ratios).
For this reason, our model will contain a superposition of three terms in which the parameters of the model are permitted to be functions of mass ratio.

It will also be convenient when we perform the modeling to scale by frequency, so that our model is calibrated on the rescaled amplitude $f|\tilde{h}_{20}|$.
This choice removes the low-frequency $f^{-1}$ behavior shared by all mass ratios, leaving a function that approaches a constant at low frequencies and decays exponentially at high frequencies.
Our model ansatz for this rescaled amplitude will be
\begin{align} \label{eq:amplitude_model}
    \frac{r}{M^2} f|\tilde{h}_{20}|= {} & A_1 M f \text{csch}\left(\frac{B_1 M f \pi}{2}\right) \nonumber \\
    & + A_2 M f \text{csch}\left(\frac{B_2 M f \pi}{2}\right)\nonumber \\
    & - A_3 (Mf)^C \text{csch}\left(\frac{B_3 M f \pi}{2}\right) \, .
\end{align}
Because the frequency $f$ has units of $M^{-1}$, we introduced factors of $M$ so that the amplitudes $A_i$ (for $i=1,2,3$), time scales $B_i$ (also for $i=1,2,3$), and power law $C$ are all dimensionless.
The factor of $\pi/2$ is conventional, and could have been absorbed into the definition of $B_i$.

To have a model that can be calibrated for mass ratios $q \in [1,8]$, we allow the model's coefficients to be functions of the symmetric mass ratio $\eta = q/(1+q)^2$.
In particular, we assume that $A_i$, $B_i$, and the power $C$ can be expressed as linear or quadratic polynomials in $\eta$:
\begin{subequations} \label{eq:amplitude_coefficients}
    \begin{align}
        A_i = {}& a_{i1} \eta + a_{i2} \eta^2 \, ,\\
        B_i = {}& b_{i0} + b_{i1} \eta \, ,\\
        C = {}& c_0 + c_1 \eta \, ,
    \end{align}
\end{subequations}
There are no constant ($\eta$ independent) terms in the amplitudes $A_i$ because the memory signal goes to zero in the limit $\eta \rightarrow 0$.
As we discuss in more detail in Sec.~\ref{subsec:calibration}, the coefficient $a_{31}$ was not strongly constrained by the calibration procedure, and we were able to set it to zero.
The fact that a two-term expression was sufficient to for the remaining coefficients  $A_i$, $B_i$, and $C$ was determined through multiple trials, and will be covered in Sec.~\ref{subsec:calibration}.

\subsection{Phenomenological phase model} \label{subsec:Phenom_model_phase}

Rather than directly model the phase of the memory waveform $\Phi(f)\equiv \arg[\tilde{h}_{20}(f)]$, we instead model the reduced phase $\hat \Phi$ defined in Eq.~\eqref{eq:hatPhiMem}.
As was described in Sec.~\ref{subsec:FD_memory_signal_properties}, the $\hat \Phi$ was defined so as to remove the $Mf \rightarrow 0$ limit of the phase (which approaches a constant $-\pi/2$) and to subtract a contribution to the phase, $2\pi t_f f$, that arises from the location of the peak of the oscillatory strain differing from the end point of the time interval.
Note that in this paper we choose the length of time after the peak to be $t_f=130M$.

The three features of this phase that we try to capture are the initial increase at very low frequencies, and the two different approximately linear slopes in two frequency ranges (for most mass ratios).
A function that behaves linearly at larger frequencies and approaches zero at lower frequencies is of the form $f e^{-f_0/f}$.
Conversely, a function of the form $f e^{-f/f_0}$ goes to zero exponentially for $f \gg f_0$ and is linear for $f \ll f_0$.
A sum of the two can capture the two different approximately linear functions.
To obtain the increase of the phase at very low frequencies, we can use a power law of the form $(Mf)^p$ for a power $p < 1$ (so that it grows more rapidly than linearly as $Mf\rightarrow 0$).

For these reasons, our ansatz for the residual phase $\hat\Phi$ will be of the form
\begin{equation} \label{eq:reduced_phase}
    \hat\Phi(f) = - \alpha Mf e^{-\beta/(Mf)} - \gamma M f e^{-Mf/\delta} + \kappa (Mf)^\lambda \, .
\end{equation}
We use Greek letters to denote the parameters of the phase model, and factors of $M$ were again introduced to make all the model parameters dimensionless.
As with the amplitude model, we will make them all the parameters functions of the symmetric mass ratio $\eta$.
We assume they are all second-order polynomials in $\eta$ of the form
\begin{equation}
\label{eq:phase_coefficients}
    x  =  x_0  +  x_1  \eta  +  x_2  \eta^2 \, ,
\end{equation}
where $x$ represents one of the six parameters in the model $(\alpha,\beta,\gamma,\delta,\kappa,\lambda)$.
For example, $\alpha$ is expanded as $\alpha = \alpha_0 + \alpha_1 \eta + \alpha_2 \eta^2 $ (and similarly for the other five parameters).

Whether the ans\"atze for the amplitude and phase models can accurately capture the properties of the memory signals as a function of mass ratio depends on whether suitable values of the parameters can be found.
The calibration procedure that we follow to obtain such parameters is given next.

\subsection{Calibrating the amplitude and phase models} \label{subsec:calibration}

\begin{table}[t]
\centering
\caption{Calibrated values of the coefficients of the frequency-domain amplitude model in Eqs.\eqref{eq:amplitude_model}--\eqref{eq:amplitude_coefficients}. 
The coefficient $a_{31}$ is set to zero so that $A_3$ depends only on $\eta^2$.}
\begin{tabular}{c c S[table-format = -1.8e-1]}
\hline\hline
Parameter & Coefficient & {Value} \\
\hline

\multirow{2}{*}{$A_1$} & $a_{11}$ & -4.62939930e-1 \\
      & $a_{12}$ & 8.05972250e-1 \\

\multirow{2}{*}{$A_2$} & $a_{21}$ & -1.93002245e0 \\
      & $a_{22}$ & -7.92143357e-1 \\

\multirow{2}{*}{$A_3$} & $a_{31}$ & {0 (fixed)} \\
      & $a_{32}$ & 1.09945379e0 \\

\multirow{2}{*}{$B_1$} & $b_{10}$ & 1.80967806e0 \\
      & $b_{11}$ & 1.97447613e-4 \\

\multirow{2}{*}{$B_2$} & $b_{20}$ & 4.64415781e0 \\
      & $b_{21}$ & 2.36292617e0 \\

\multirow{2}{*}{$B_3$} & $b_{30}$ & 1.82272214e0 \\
      & $b_{31}$ & 1.61541350e-5 \\

\multirow{2}{*}{$C$}   & $c_{0}$ & 9.99923604e-1 \\
      & $c_{1}$ & 9.49905432e-1 \\

\hline\hline
\end{tabular}
\label{tab:amp_coefficients}
\end{table}

\begin{table*}[t]
\centering
\caption{Calibrated coefficients of the frequency-domain phase model in Eqs.~\eqref{eq:reduced_phase}--\eqref{eq:phase_coefficients}. 
Each parameter is modeled as $x=x_0+x_1\eta+x_2\eta^2$ for $x=\alpha,\beta,\gamma,\delta,\kappa,\lambda$.}
\begin{tabular}{c S[table-format = -1.8e-1] S[table-format = -1.8e-1] S[table-format = -1.8e-1]}
\hline\hline
Parameter & {$x_0$} & {$x_1$} & {$x_2$} \\
\hline

$\alpha$ &
1.54644804e2 &
-3.83187952e2 &
-1.09433432e2 \\

$\beta$ &
-4.53639279e-2 &
1.96682245e0 &
-6.42498632e0 \\

$\gamma$ &
-4.90466459e0 &
4.66409968e2 &
5.54509756e2 \\

$\delta$ &
-4.90706383e-2 &
1.42544528e0 &
-2.91977247e0 \\

$\kappa$ &
6.54194457e0 &
-1.30731611e2 &
7.91180733e2 \\

$\lambda$ &
-2.92996886e-1 &
6.37110873e0 &
-1.00876756e1 \\

\hline\hline
\end{tabular}
\label{tab:phase_coefficients}
\end{table*}

The values of the calibrated coefficients for the amplitude model are given in in Table~\ref{tab:amp_coefficients} and those of the phase model are given in Table~\ref{tab:phase_coefficients}.
We now describe how we obtained these coefficients.

First, we compute the GW memory signal from the surrogate model at eight mass ratios $q=1,\ldots,8$.
The surrogate is evaluated at a uniformly spaced set of times (mentioned in Sec.~\ref{subsec:FFT}), which is mapped to a uniformly space set of frequencies when the FFT is performed.
To ensure that the low-frequency behavior of the memory signal is properly resolved, we interpolate the memory signal computed from NR surrogate to a set of frequencies that are spaced uniformly in the logarithm of frequency.
The interpolation is done with a cubic splines implemented in \textsc{SciPy}~\cite{2020SciPy-NMeth}.

The model parameters are determined by performing a global simultaneous fit for the eight mass ratios $q=1,\ldots,8$, similar to the calibration procedure used for the time-domain model of Paper~II.
We minimize a cost function defined from the superposition of eight cost functions at each mass ratio $q$,
\begin{equation} \label{eq:cost_function}
    C[g^{\surr},g^{\model}] =  \sum_{q=1}^{8} C_q[g^{\surr},g^{\model}] \, ,
\end{equation}
where the cost function for an individual mass ratio, $C_q[g^{\surr},g^{\model}]$, is defined by 
\begin{equation} \label{eq:cost_function_q}
    C_q[g_{(A)},g_{(B)}] \equiv \frac{\displaystyle\int_{f_\mathrm{min}}^{f_\mathrm{max}} df \, |g^{(A)}(f;q)-g^{(B)}(f;q)|^2}{\displaystyle\int_{f_\mathrm{min}}^{f_\mathrm{max}} df \, |g^{(A)}(f;q)|^2} \, .
\end{equation}
Here $g$ represents either the residual phase $\hat \Phi(f)$ or the rescaled amplitude $f r|\tilde h_{20}(f)|/M$ of the GW memory signal.
The integrals run from $f_{\rm min}=10^{-6}$ (set by the length of the surrogate memory signal) to $f_{\rm max} = 0.08\,M^{-1}$, which we noted above was the frequency above which the surrogate amplitude has noise artifacts (i.e., it begins deviating from the corresponding NR data on which it was calibrated).
The total cost function is a linear superposition of cost functions at each mass ratio, which assigns equal weight to all of the mass ratios used in the training.

We used the \textsc{SciPy} \emph{minimize} function to obtain the values for the 14 parameters in the amplitude model in Table~\ref{tab:amp_coefficients}.
When the minimize function searched over different parameter ranges, we found that the value of $a_{31}$ was not constrained strongly by the optimization of the cost function and could be set to zero without causing any significant change in the value of the cost function.
This is why we are able to write $A_3 = a_{32}\eta^2$ as a monomial rather than a polynomial in $\eta$.
For the phase model, we followed the same procedure as we did for the amplitude model to now obtain the values of the 18 parameters listed in Table~\ref{tab:phase_coefficients}.
In this case, we found that all of the coefficients had an impact on the cost function, which is why all values are nonzero in the Table.

Both the amplitude and phase models have been implemented in the \texttt{GWMemoryModel}~\cite{code:GWMemoryModel} (which also has an implementation of the time-domain model of Paper~II).

\section{Results and comparisons} \label{sec:results}

We present the memory signals generated by our model in this section.
We compare them with the surrogate model from which they were calibrated and the time-domain model from Paper~II.
Finally, we compute the mismatch over the parameter space of the model.

\subsection{Comparison with the memory signal computed from the surrogate model}

\begin{figure*}
    \centering
    \includegraphics[width=\columnwidth]{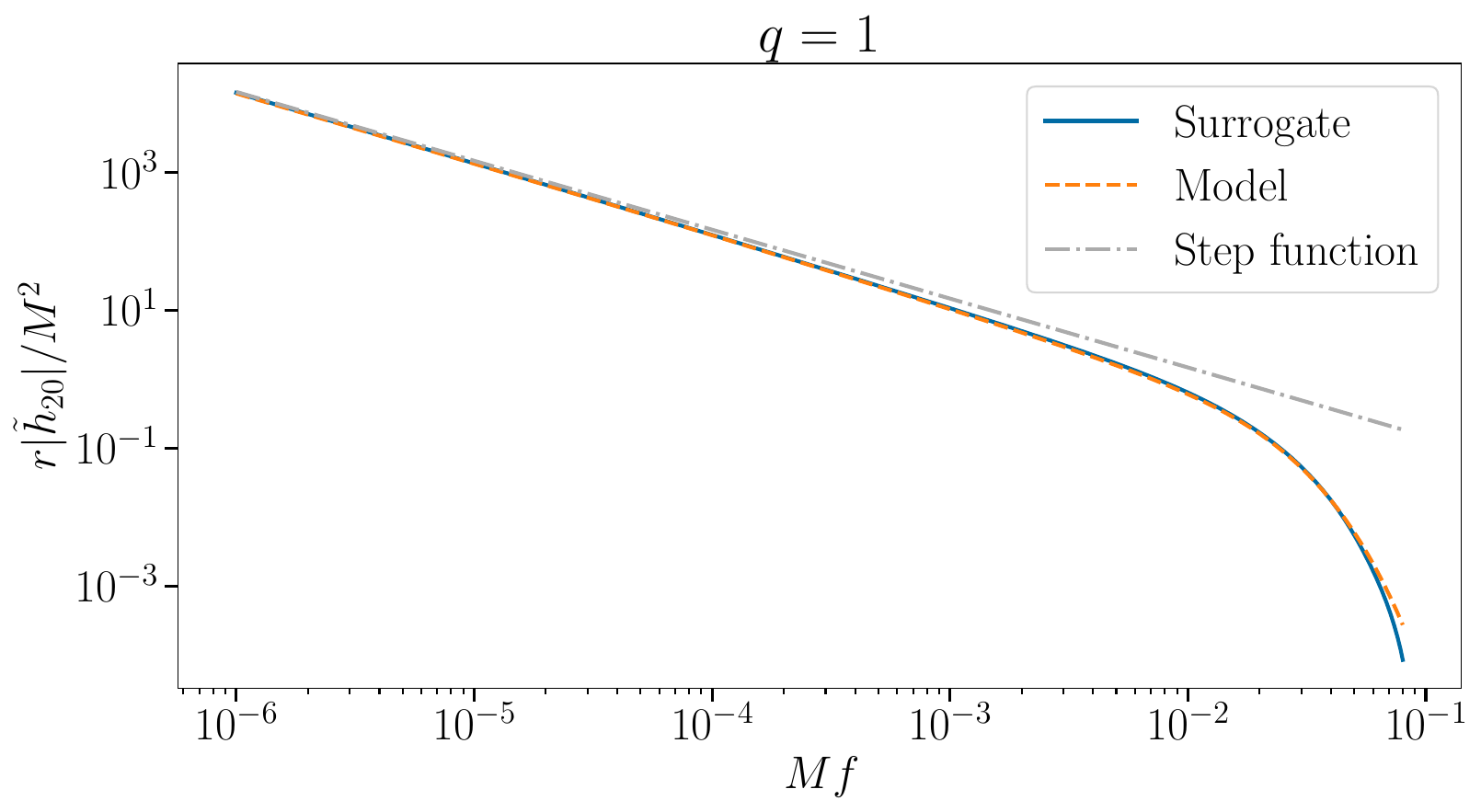}
    \includegraphics[width=\columnwidth]{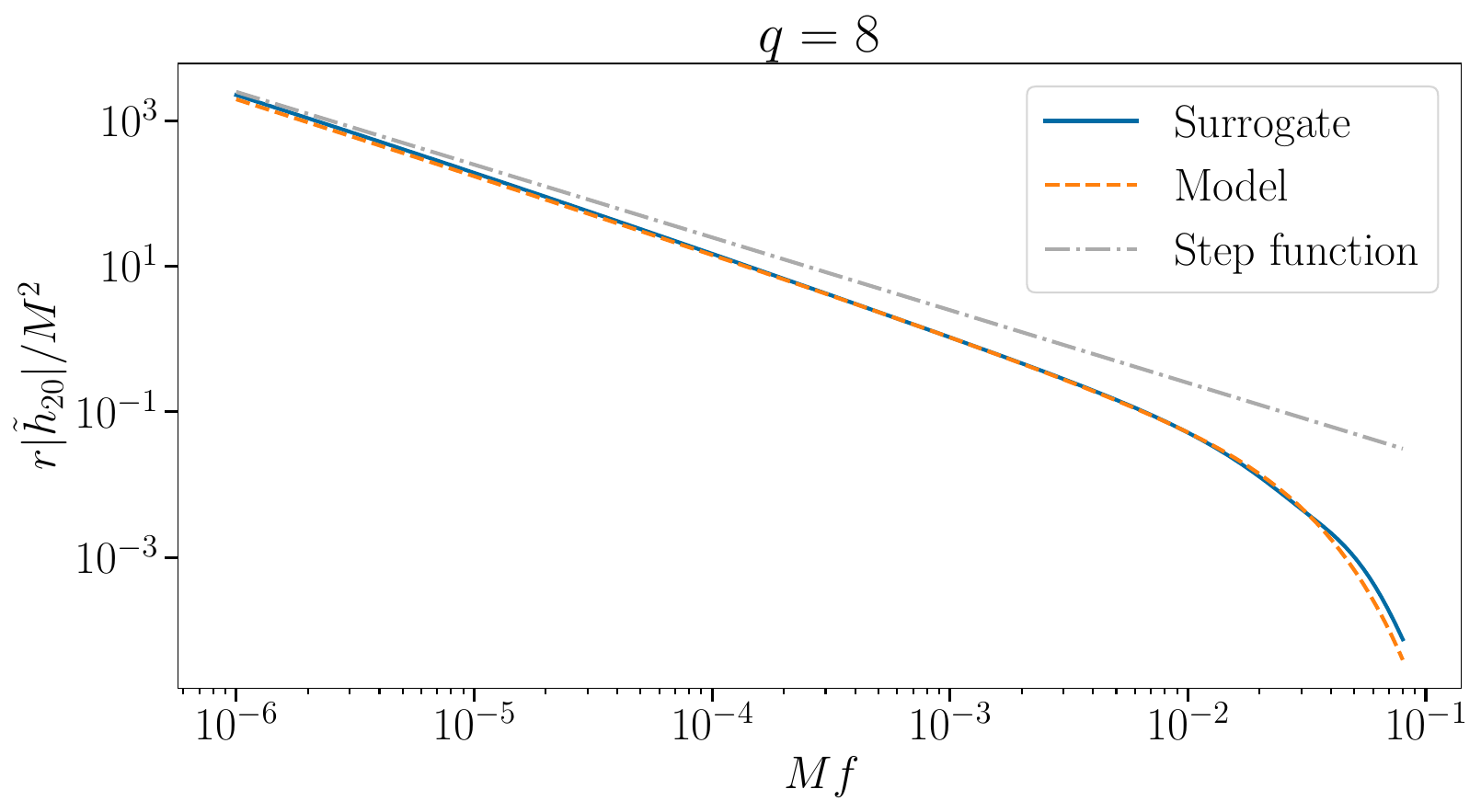}
    \includegraphics[width=\columnwidth]{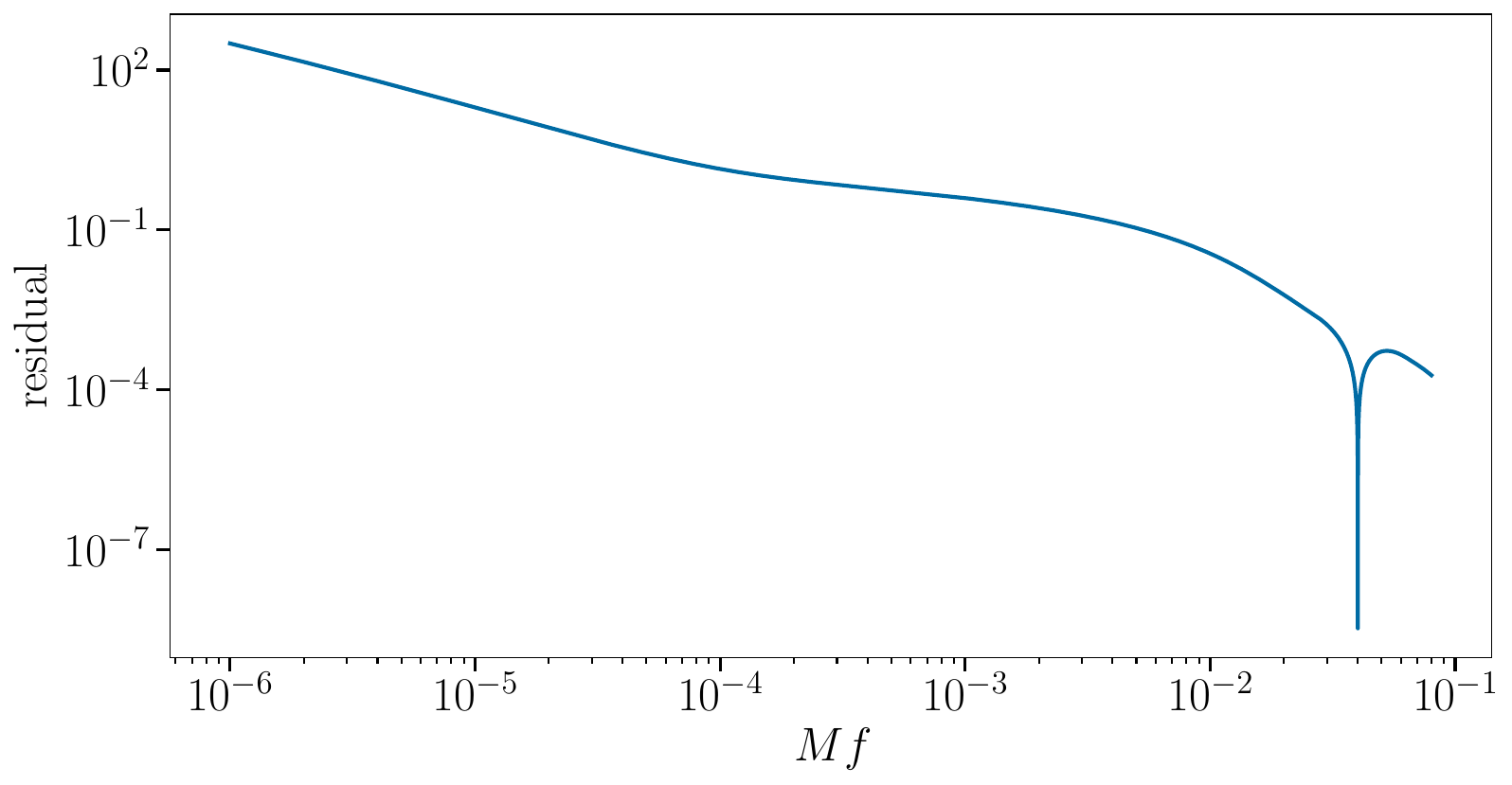}
    \includegraphics[width=\columnwidth]{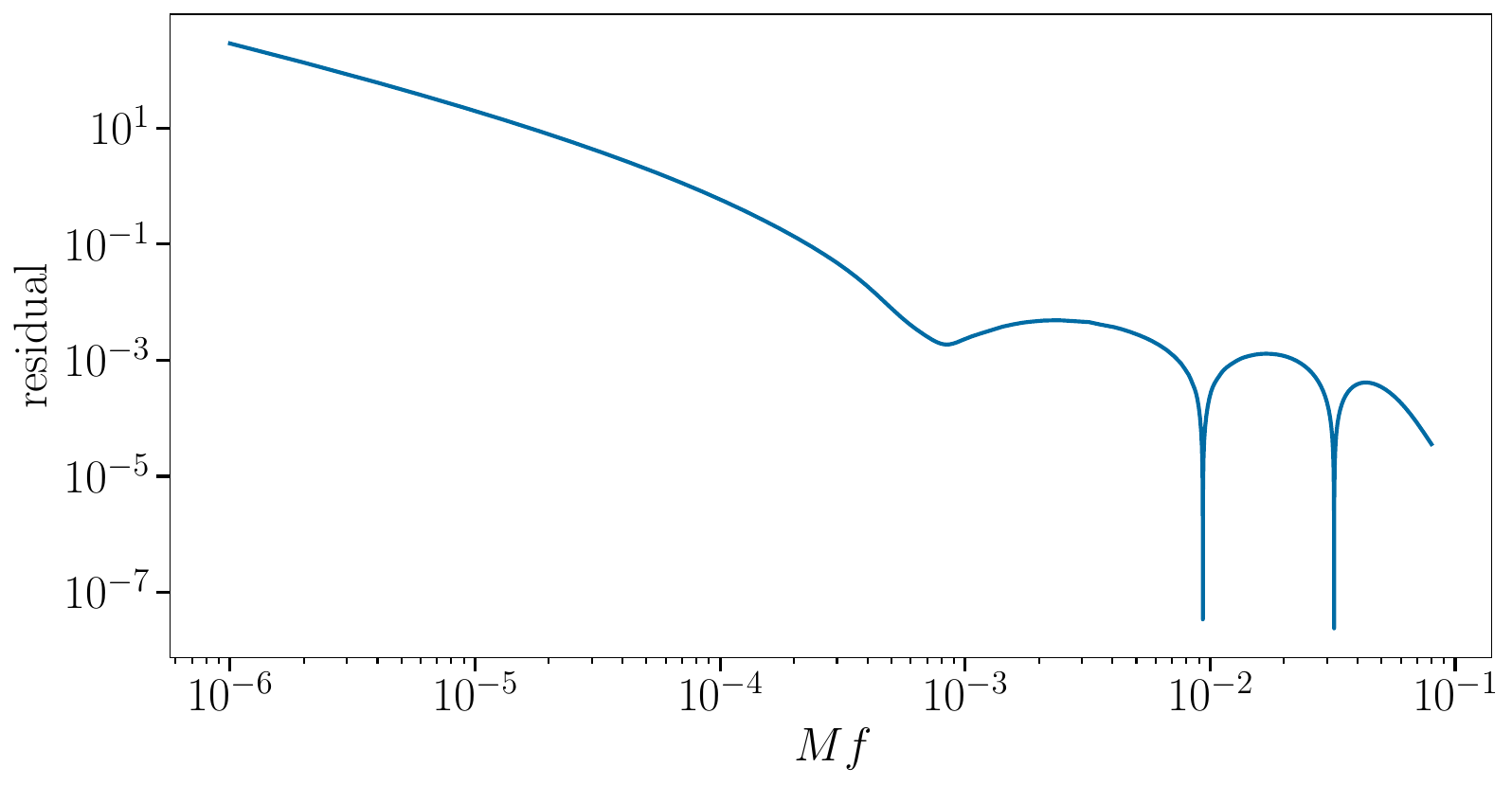}
    \caption{\textbf{Amplitude of the frequency-domain GW memory signal versus frequency}: 
    \emph{Top}: The amplitude $r|\tilde h_{20}(f)|/M^2$ of the surrogate memory signal (solid blue), the phenomenological, frequency-domain amplitude model (dashed orange), and the step-function approximation of the memory signal (dash-dotted gray) are shown as functions of the frequency $Mf$ for the smallest and largest mass ratios covered by our model ($q=1$ on the left and $q=8$ on the right).
    \emph{Bottom}: The residuals (magnitude of the difference) between the frequency-domain amplitude phase model and the memory signal calculation from the surrogate model.}
    \label{fig:memory_amp}
\end{figure*}

We show the amplitude model $|r \tilde{h}_{20}^{\rm model}/M^2|$ as a function of frequency for the mass ratios $q=1,8$ in Fig.~\ref{fig:memory_amp}.
In the top panels, our frequency-domain model is the dashed orange curve, the step function approximation is the dash-dotted gray curve, and the GW memory computed from the surrogate model is the solid blue curve.
The mass ratio $q=1$ is on the left and $q=8$ is on the right.
The model accurately reproduces the surrogate amplitude (solid blue curve) across the full frequency range for both mass ratios.
This is shown more quantitatively in the bottom panels of Fig.~\ref{fig:memory_amp}.
They show the residuals $|r(|\tilde{h}_{20}^{\surr}| - |\tilde{h}_{20}^{\model}|)/M^2|$ between the amplitude model and the surrogate memory signal for each mass ratio.
The residual decreases from $\sim 10^2$ at the lowest frequencies to $\sim10^{-4}$ near $M f\sim 10^{-2}$ in both panels.
However, the relative error between the surrogate calculation and the frequency-domain model grows somewhat at the higher frequencies given the exponential decay in the amplitudes of both the model and surrogate-based calculation.

\begin{figure*}
    \centering
    \includegraphics[width=\columnwidth]{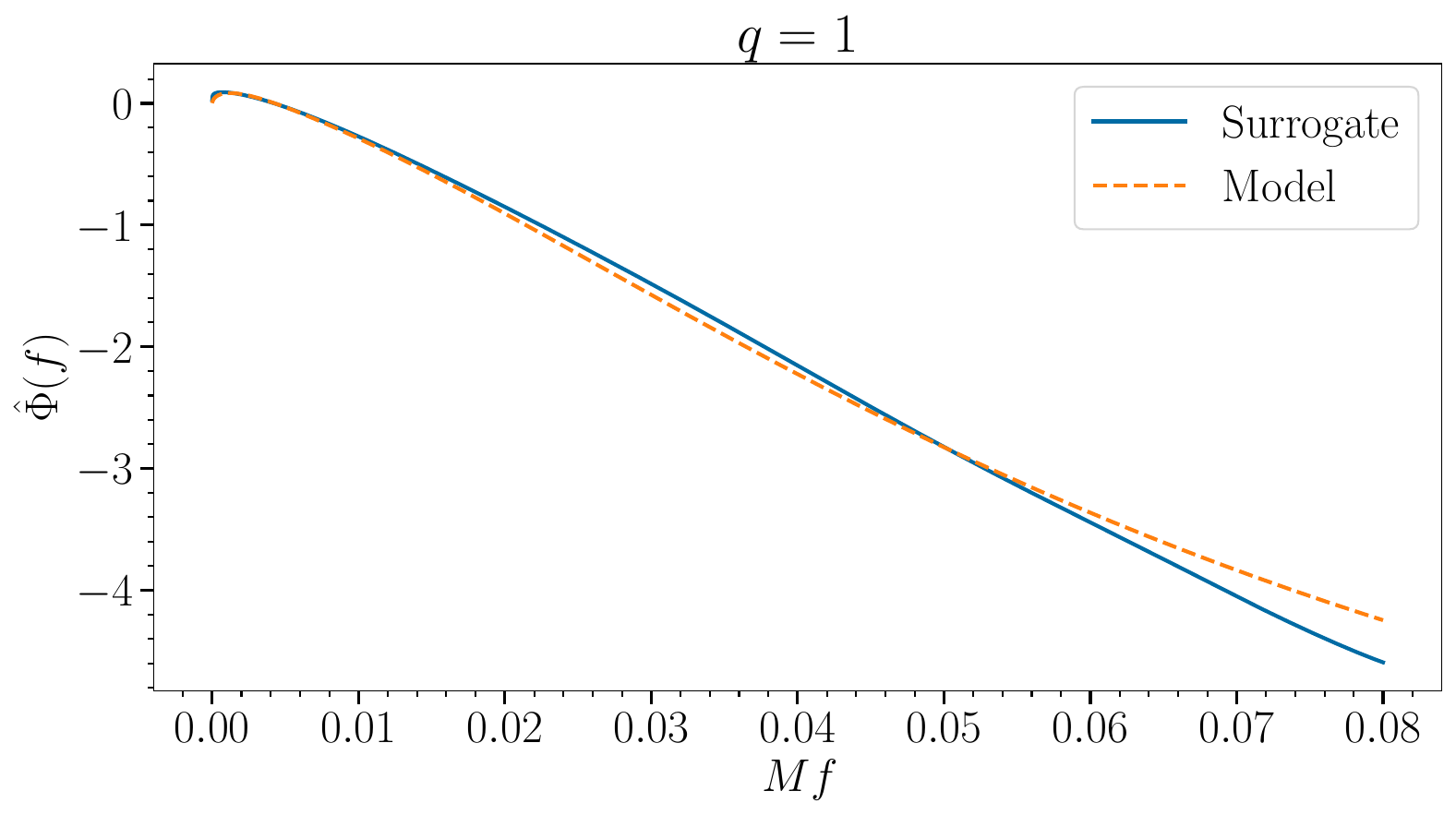}
    \includegraphics[width=\columnwidth]{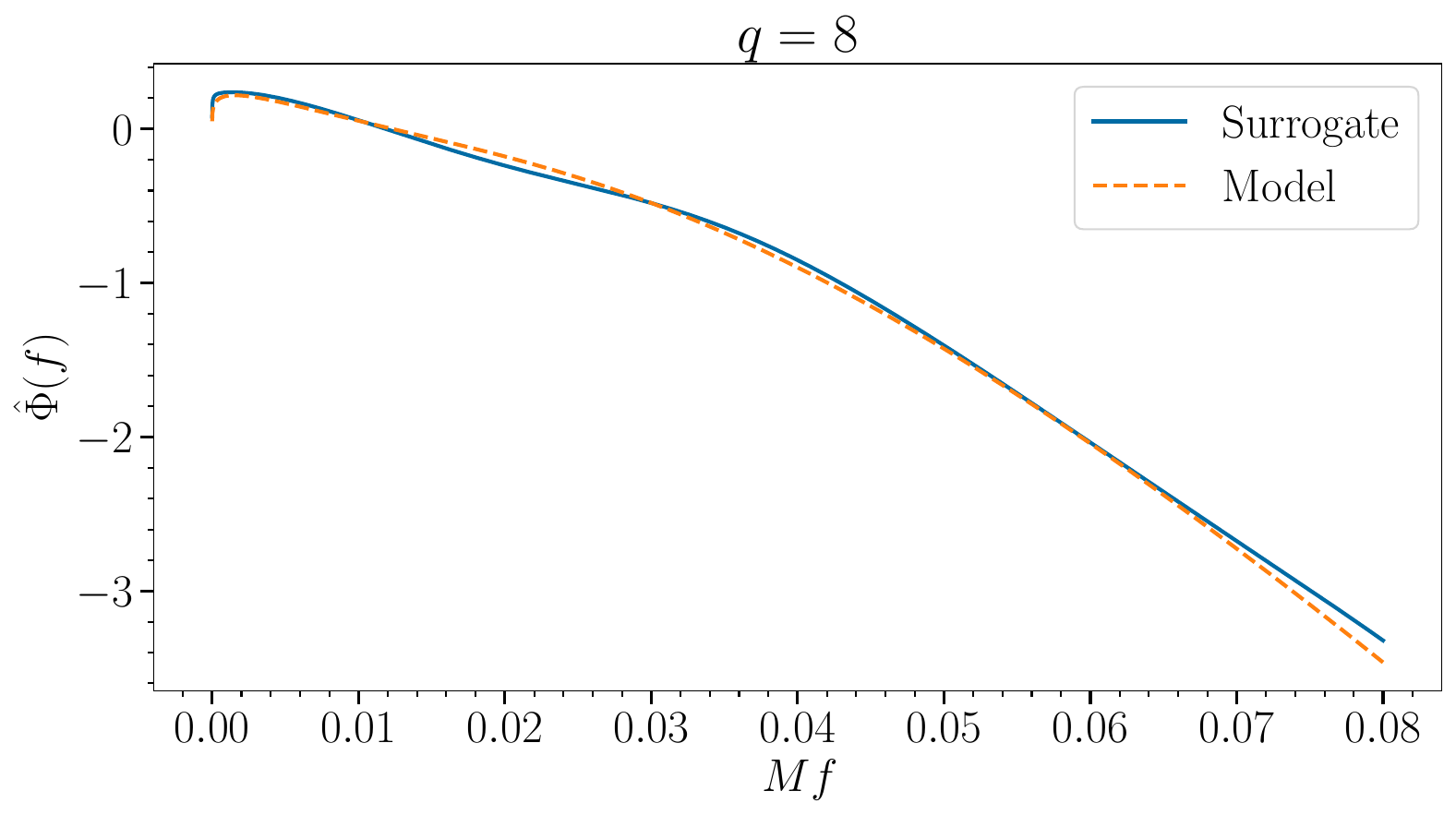}
    \includegraphics[width=\columnwidth]{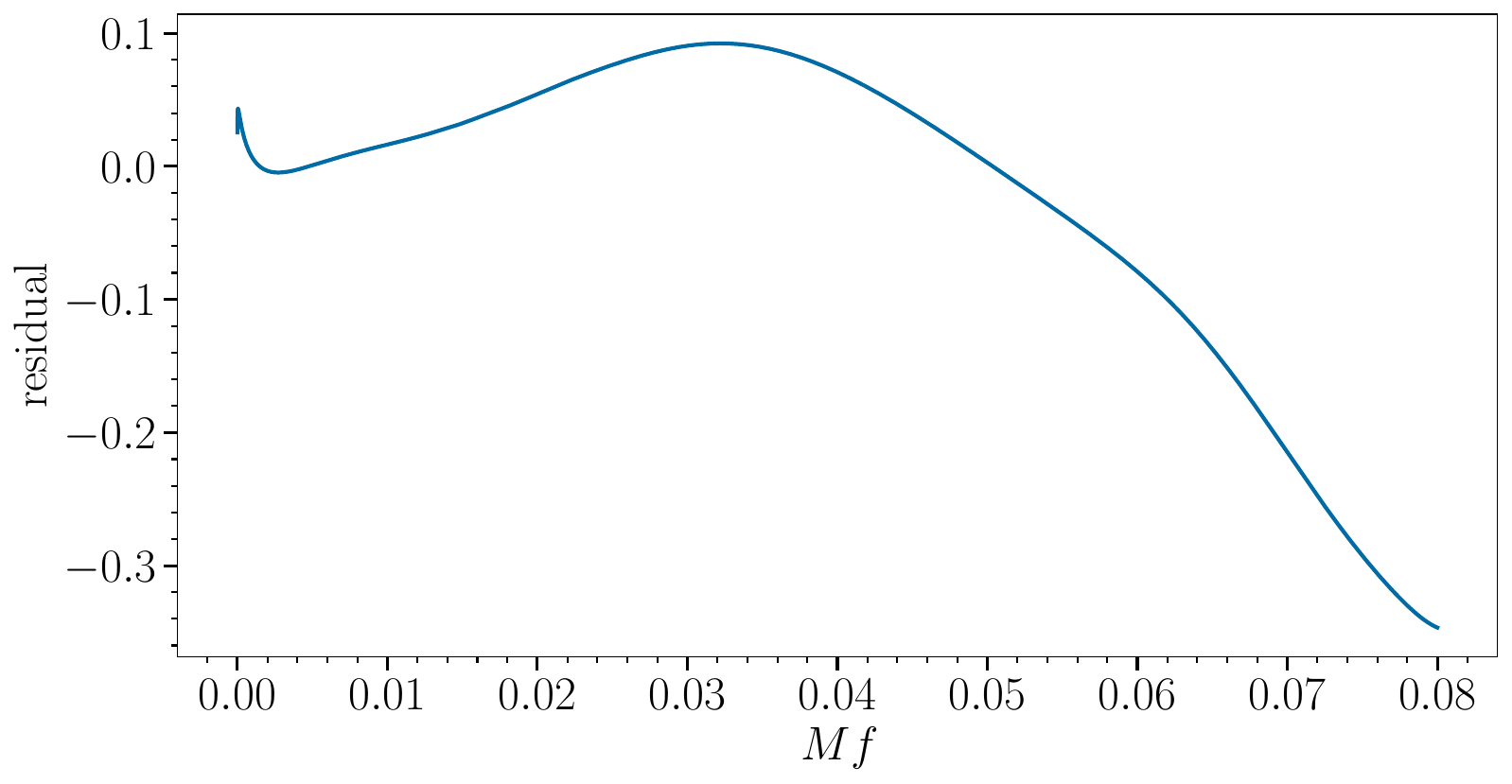}
    \includegraphics[width=\columnwidth]{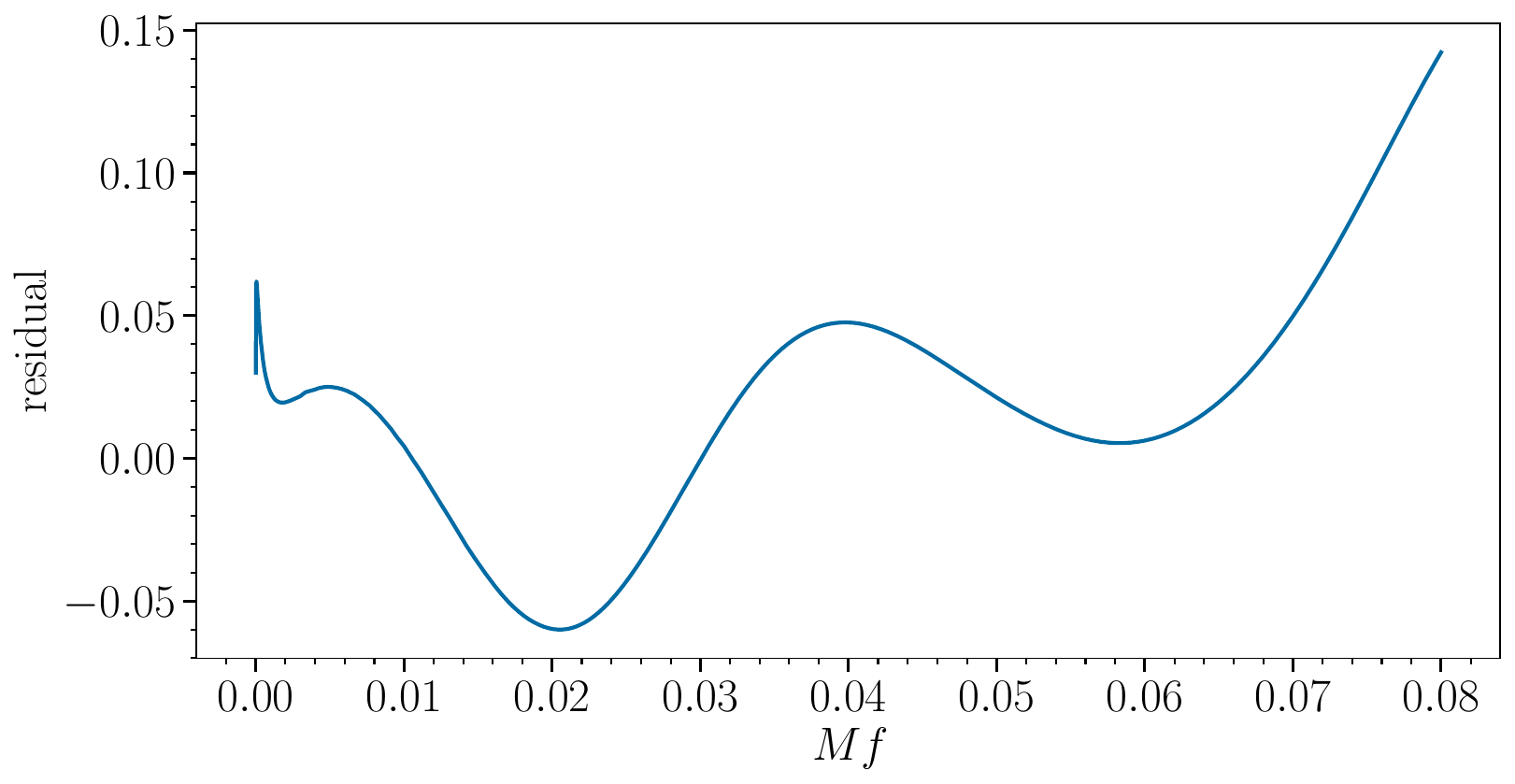}
    \caption{\textbf{Phase of the frequency-domain GW memory signal versus frequency}: 
    \emph{Top}: The rescaled phase $\hat\Phi(f) \equiv \arg[\tilde h_{20}(f)] +\pi/2 - 2 \pi t_f f$ of the surrogate memory signal (solid blue) and the phenomenological frequency-domain phase model (dashed orange) are shown as a function of frequency $Mf$ for mass ratios $q = 1$ and $q = 8$ in the left and right panels, respectively.
    \emph{Bottom}: The residual between the frequency-domain phase model and the phase of the corresponding result computed from the NR surrogate.}
    \label{fig:memory_phase}
\end{figure*}

In the top panels of Fig.~\ref{fig:memory_phase}, the frequency-domain phenomenological phase model for $\hat \Phi$ is plotted as an orange dashed curve and the corresponding phase computed from the surrogate model is shown as the solid blue curve.
Again, the mass ratio $q=1$ is on the left and $q=8$ is on the right.
The residuals shown in the bottom panels are of order $10^{-1}$, which corresponds to a similar relative error given that the phase is of order one.
While the phase errors are largest at the highest frequencies shown in Fig.~\ref{fig:memory_phase}, it is useful to keep in mind that the overall amplitude of the signal is exponentially suppressed in this region.
Thus, for calculations with GW detectors, the impact of this larger phase error may not be as significant (depending on the total mass of the system).
A more quantitative assessment will be made in Sec.~\ref{subsec:mismatch}, where the mismatch is computed.

\subsection{Comparison with the time-domain model of Paper II}
\label{subsec:models_comparison}

We perform a few comparisons of the new frequency-domain GW memory signal model in this paper with the time-domain model of Paper~II.
To make this comparison, we perform the FFT of the time-domain memory model using the same approach described in Sec.~\ref{subsec:FFT}, except we compute the time-domain model over a time of order $10^4 M$ rather than $10^6 M$.

\begin{figure}
    \centering
    \includegraphics[width=\columnwidth]{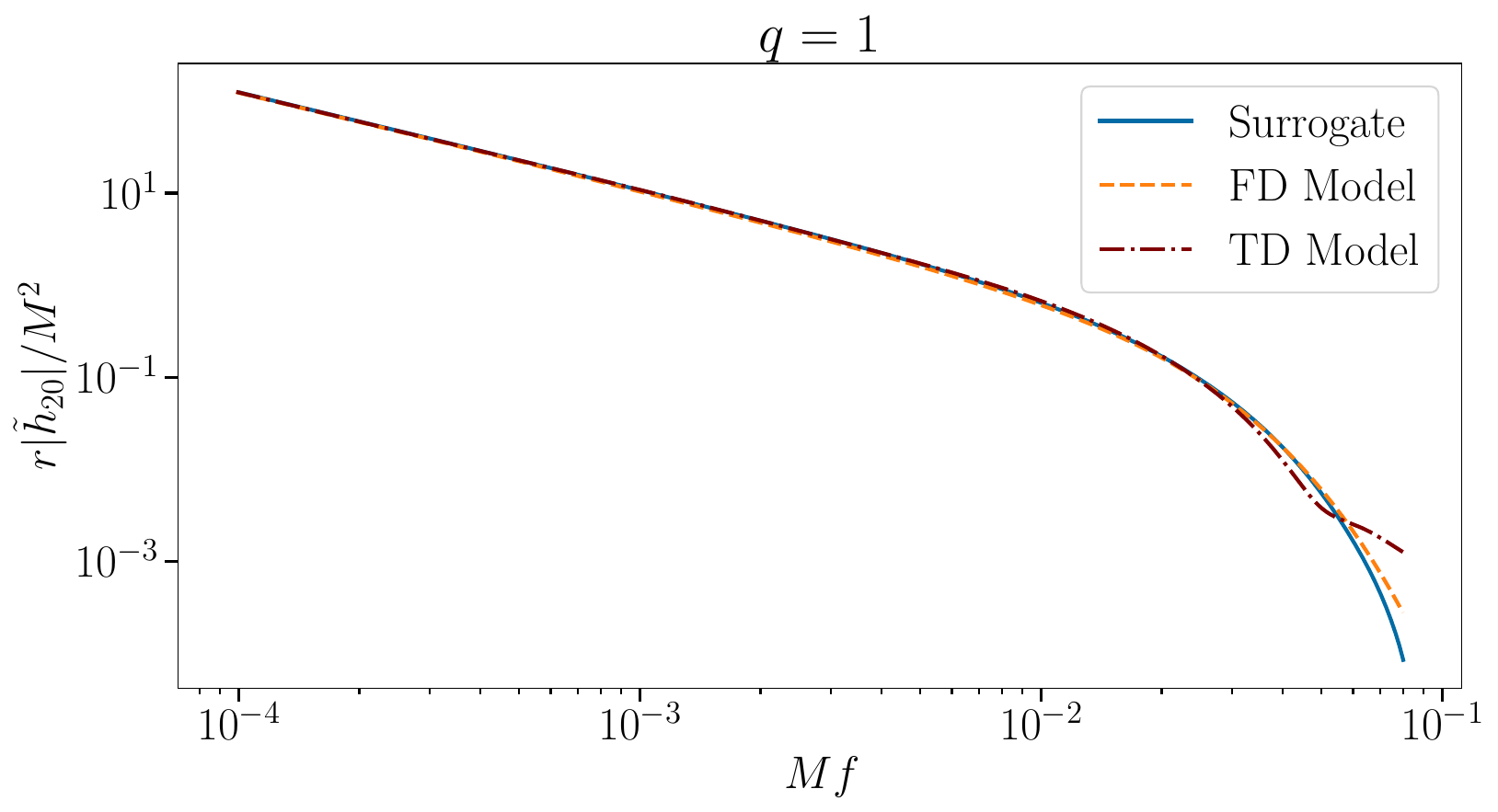}
    \includegraphics[width=\columnwidth]{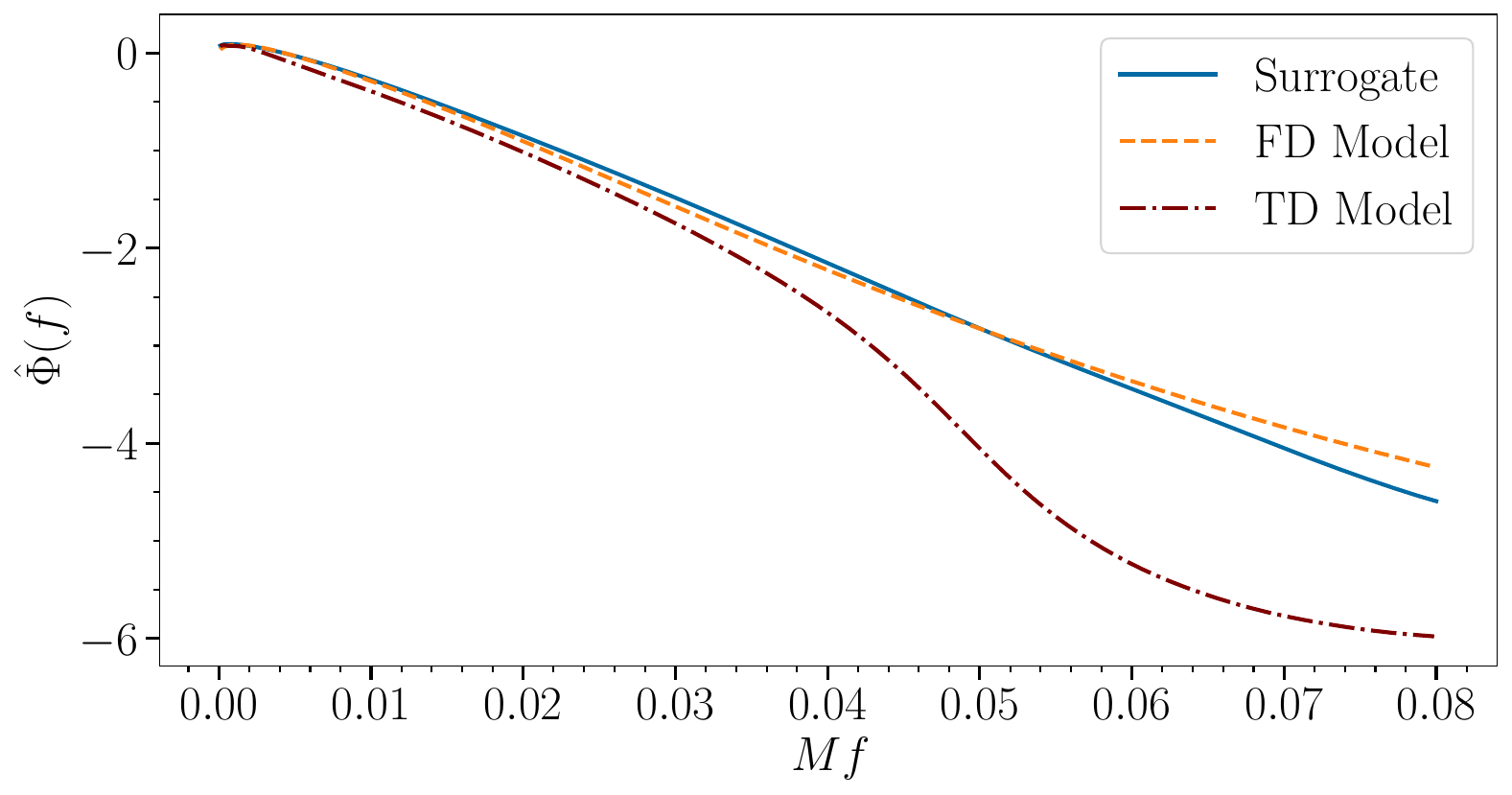}
    \caption{\textbf{Comparison between the phenomenological frequency-domain (FD) model, the Fourier transform of the time-domain (TD) model, and the surrogate memory signal for an equal-mass binary.}
    \emph{Top}: The amplitude $r|\tilde h_{20}(f)|/M^2$ as a function of the frequency $Mf$. The FD model (dashed orange curve) closely reproduces the surrogate amplitude (solid blue curve) more closely over the frequencies shown.
    The FFT of the TD model (dash-dotted maroon curve) deviates more at the highest frequencies because the third derivative is discontinuous at two times (as reviewed in more detail in Sec.~\ref{subsec:propsFT}).
    \emph{Bottom}: The residual phase $\hat\Phi(f)$ of the same signals. 
    The FD model more closely matches the frequency dependence of the surrogate-based calculation's phase, whereas the FFT of the TD model approaches $-2\pi$ at higher frequencies, which is related to the discontinuities in the third time derivative of the model (see Sec.~\ref{subsec:propsFT}).}
    \label{fig:models_camparisons}
\end{figure}

Figure~\ref{fig:models_camparisons} shows the amplitude $r|\tilde{h}_{20}(f)|/M^2$ and reduced phase $\hat\Phi(f)$ for an equal-mass binary ($q=1$).
It shows three cases, the GW memory calculated from the surrogate (the solid blue curves), the phenomenological frequency-domain model (the orange dashed curves labeled by FD), and the FFT of the time-domain model (the dash-dotted maroon curves labeled by TD).
In the top panel, the amplitude of the phenomenological model agrees more closely with the surrogate calculation of the GW memory signal than the FFT of the time-domain model does.
The deviation from the exponential falloff of the time-domain model is related to the discontinuities in the third derivative of the model at the times $t_\intr$ and $t_\rd$, which was reviewed in more detail in Sec.~\ref{subsec:propsFT}.

The bottom panel illustrates the corresponding behavior of the residual phase $\hat \Phi$.
The styling of the different curves is the same as in the top panel.
For the FFT of the time-domain model, the phase approaches $-2\pi$ at high frequencies, which again occurs because of the discontinuity in the third derivative of the time-domain model at $t_\intr$ and $t_\rd$.
The phenomenological frequency-domain model is a smooth function for $f>0$ by construction; its reduced phase, thus better matches the corresponding phase from the calculation using the surrogate model.
The differences in both amplitude and phase will have an impact on the calculation of the mismatch for the most massive binaries where the higher-frequency portion of the signal is most relevant.

In addition to the more accurate representation of the GW memory signal, the new frequency-domain model is significantly faster to evaluate than the time-domain model.
Specifically, for the time-domain model we evaluate the model in the time domain and then take its FFT.
This was shown to be faster than using the analytical Fourier transform of the time-domain model in Paper~II.
We performed the timing at several mass ratios in the range $1\leq q \leq 8$.
The frequency-domain model speeds up the evaluation time by a factor of 25--35 compared to evaluating the time-domain memory signal and taking its FFT.
These timing results are also for evaluating the frequency-domain model at the same uniformly spaced frequencies that arise after taking the FFT of the time-domain model.
A further speed up in timing could be made for the frequency-domain model by evaluating it at a smaller set of irregularly spaced frequencies that capture where the signal changes most rapidly in the frequency domain.

\subsection{Accuracy of the model}
\label{subsec:mismatch}

Similar to what was done in Paper II, we quantify the accuracy of the phenomenological frequency-domain model by computing its mismatch with the FFT of the GW memory signal computed from the surrogate model.
The mismatch between two signals $h_{\surr}$ and $h_{\model}$ is defined as
\begin{equation}
\label{eq:mismatch}
    \mathcal{M} = 1-\frac{\big<h_{\surr},h_{\model}\big>}{\sqrt{\big<h_{\surr},h_{\surr}\big>\big<h_{\model},h_{\model}\big>}} ,
\end{equation}
where $\big<h_{1},h_{2}\big>$ denotes the noise-weighted inner product between two signals $h_1$ and $h_2$:
\begin{equation}\label{eq:inner_product}
    \big<h_{1},h_{2}\big> = 4\mathcal{R}\Big[\int_{f_\mathrm{min}}^{f_\mathrm{max}} \mathrm df \frac{\tilde{h}_1(f) \overline{\tilde{h}_2(f)}}{S_n(f)} \Big] \, .
\end{equation}
The quantity $S_n(f)$ in the denominator is the noise power-spectral density.
As in Paper~II, we use the Advanced LIGO design sensitivity curve for the fourth observing run (O4), which can be accessed from~\cite{LIGOpsds}.
We compute the memory signal for binaries with primary mass in the typical range of LIGO BBH detections $m_1 \in [5,100] M_\odot$ and with the secondary mass set by $m_2=m_1/q$.
The mass ratio $q$ will be restricted to the calibration range of the model $q\in[1,8]$.

\begin{figure}
    \centering
    \includegraphics[width=\columnwidth]{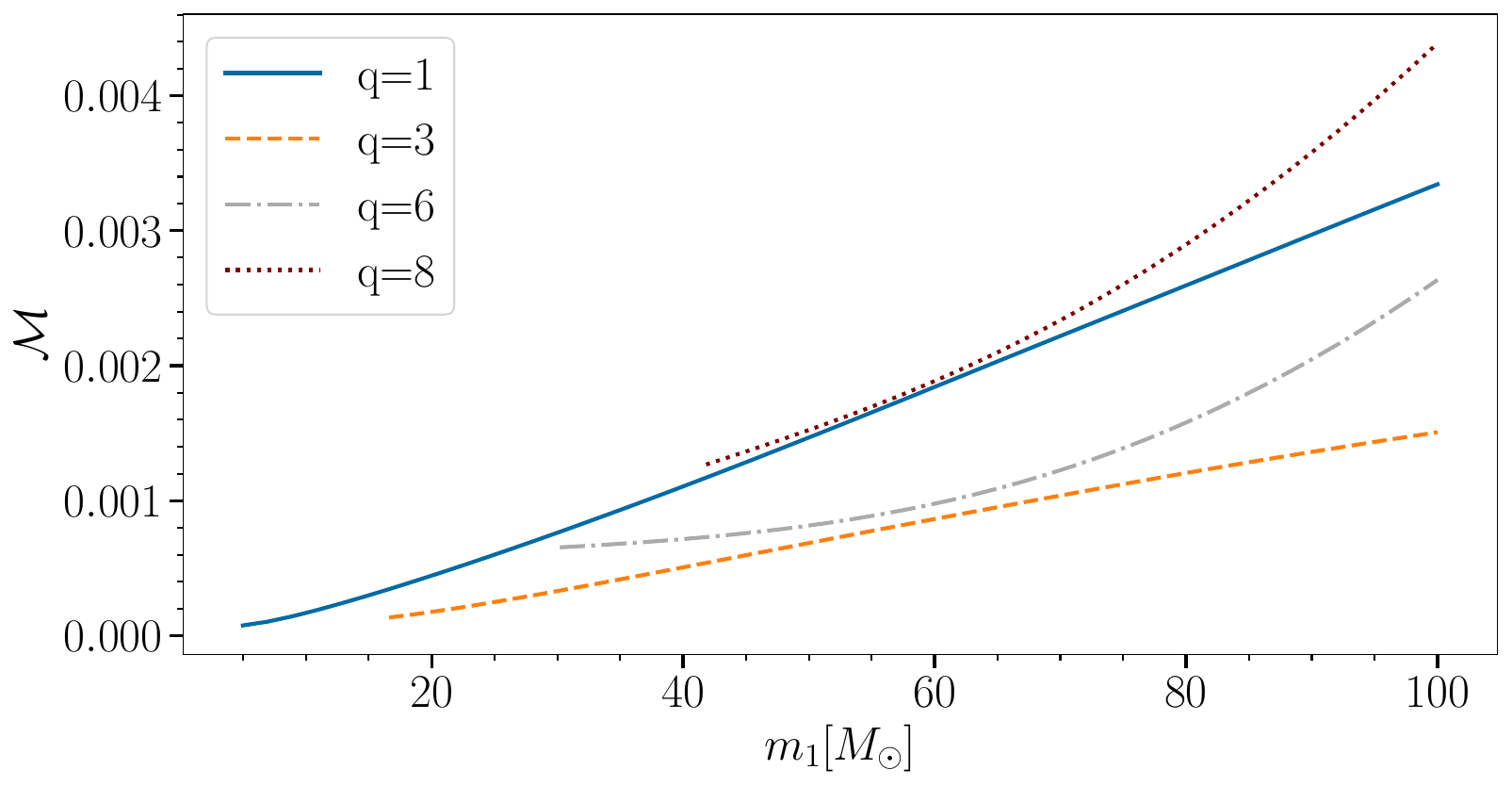}
    \includegraphics[width=\columnwidth]{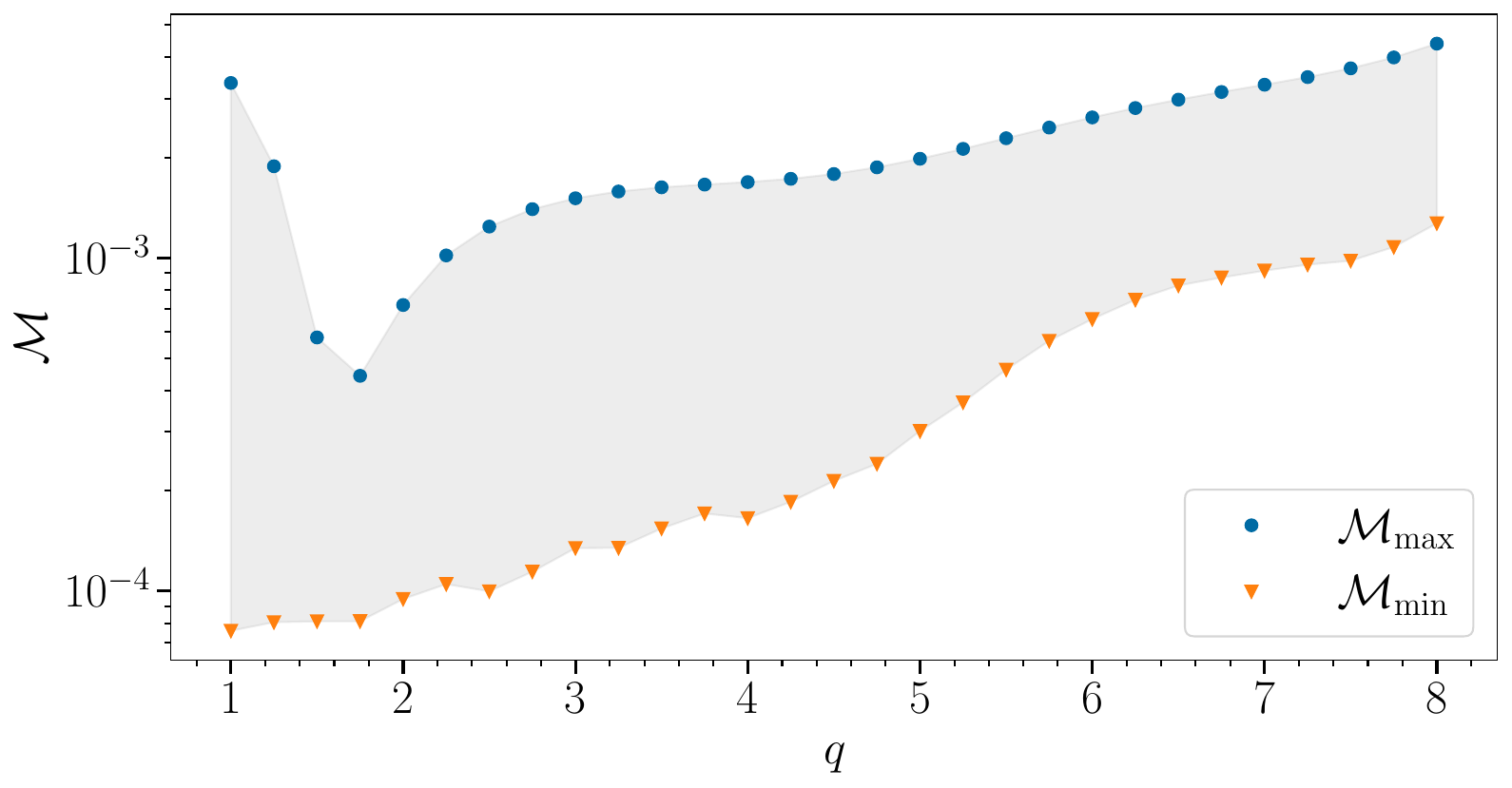}
    \caption{\textbf{Mismatch between the frequency-domain memory waveform model and the memory signal computed from the NR surrogate}.
    The Advanced LIGO design-sensitivity noise curve was used in the computation of the mismatch.
    \emph{Top}: $\mathcal{M}$ in Eq.~\eqref{eq:mismatch} a function of primary mass $m_1 \in [5\,M_\odot,100\,M_\odot]$ for mass ratios $q = 1, 3, 6, 8$, with $m_2 = m_1/q$ constrained to the same range.
    The line styles for each mass ratio are indicated in the legend.
    \emph{Bottom}: The maximum (blue circles), minimum (orange triangles), and range (shaded gray region) of mismatch values at $q$ for the primary masses $m_1 \in [5\,M_\odot,100\,M_\odot]$.
    The mismatch remains below $\sim 4 \times 10^{-3}$ across the full parameter space illustrated.
    This significantly improves the similar results for the model in Paper~II}
    \label{fig:mismatch}
\end{figure}

In the top panel of Fig.~\ref{fig:mismatch} is the mismatch between the phenomenological frequency-domain model and the memory signal computed from the surrogate is shown as a function of the primary mass $m_1$ for the mass ratios of $q=1, 3, 6, 8$ (the four different colors and line styles are summarized in the legend).
The mismatch is an increasing function of the primary mass $m_1$ for the mass ratios show, but it remains below $\mathcal{M}\sim 4\times10^{-3}$ for the mass ratios shown.
This is almost an order-of-magnitude improvement over the time-domain model developed in Paper~II.
The larger error at larger $m_1$ values is related to the fact that less of the low-frequency $1/f$ part of the memory signal within the LIGO band, where the phenomenological model and surrogate model agree most closely (and conversely, this is why the low-mass binaries have smaller mismatches).

The bottom panel of Fig.~\ref{fig:mismatch} shows the maximum mismatch (blue circles) and minimum mismatch (orange triangles) over the range of $m_1$ for each mass ratio $q$.
The variation in $m_1$ is indicated by the gray shaded region at each $q$.
As in the top panel, the larger mismatches correspond to the higher-mass systems, whereas the smaller ones are the lower-mass systems.
However, this panel gives a more complete depiction of the range of mismatch values over the entire range of masses and mass ratios.

Because the mismatch is computed from a normalized inner product between two GW signals, it primarily is a measure of the signals' agreement in phase.
To better assess the quality of the model's amplitude, we use the ``SNR mismatch'' that was introduced in Paper~II.
It was defined to be the normalized difference between the SNRs of the surrogate and model waveforms,
\begin{equation}
\label{eq:amplitude_mismatch}
    \mathcal{M}_{\rho} = \frac{\rho_{\surr}-\rho_{\model}}{\rho_{\surr}} \, .
\end{equation}
Here $\rho=\sqrt{\left<h,h\right>}$ is the optimal SNR of a waveform, which is computed using the noise-weighted inner product as in Eq.~\eqref{eq:inner_product}.

\begin{figure}
    \centering
    \includegraphics[width=0.48\textwidth]{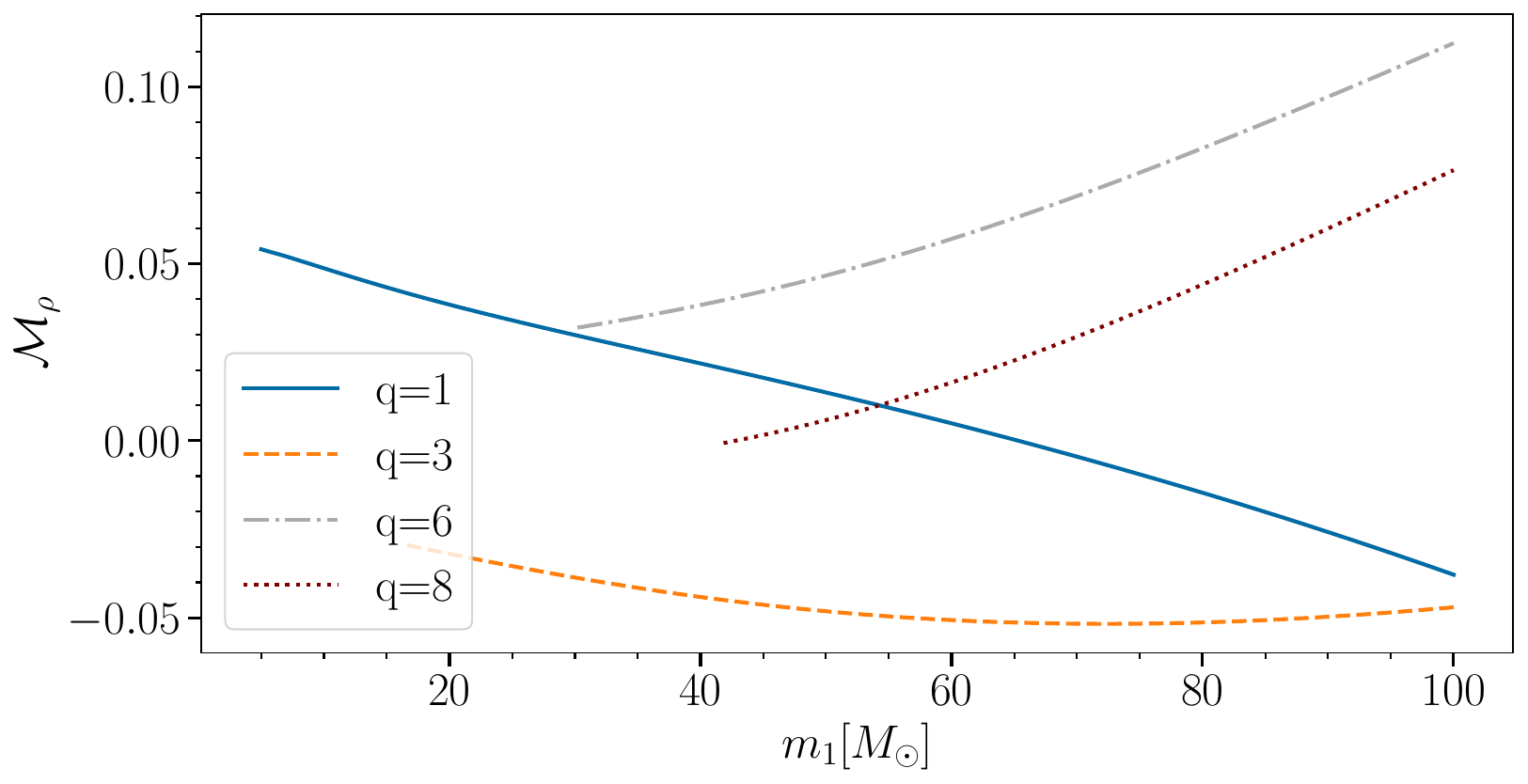}
    \includegraphics[width=0.48\textwidth]{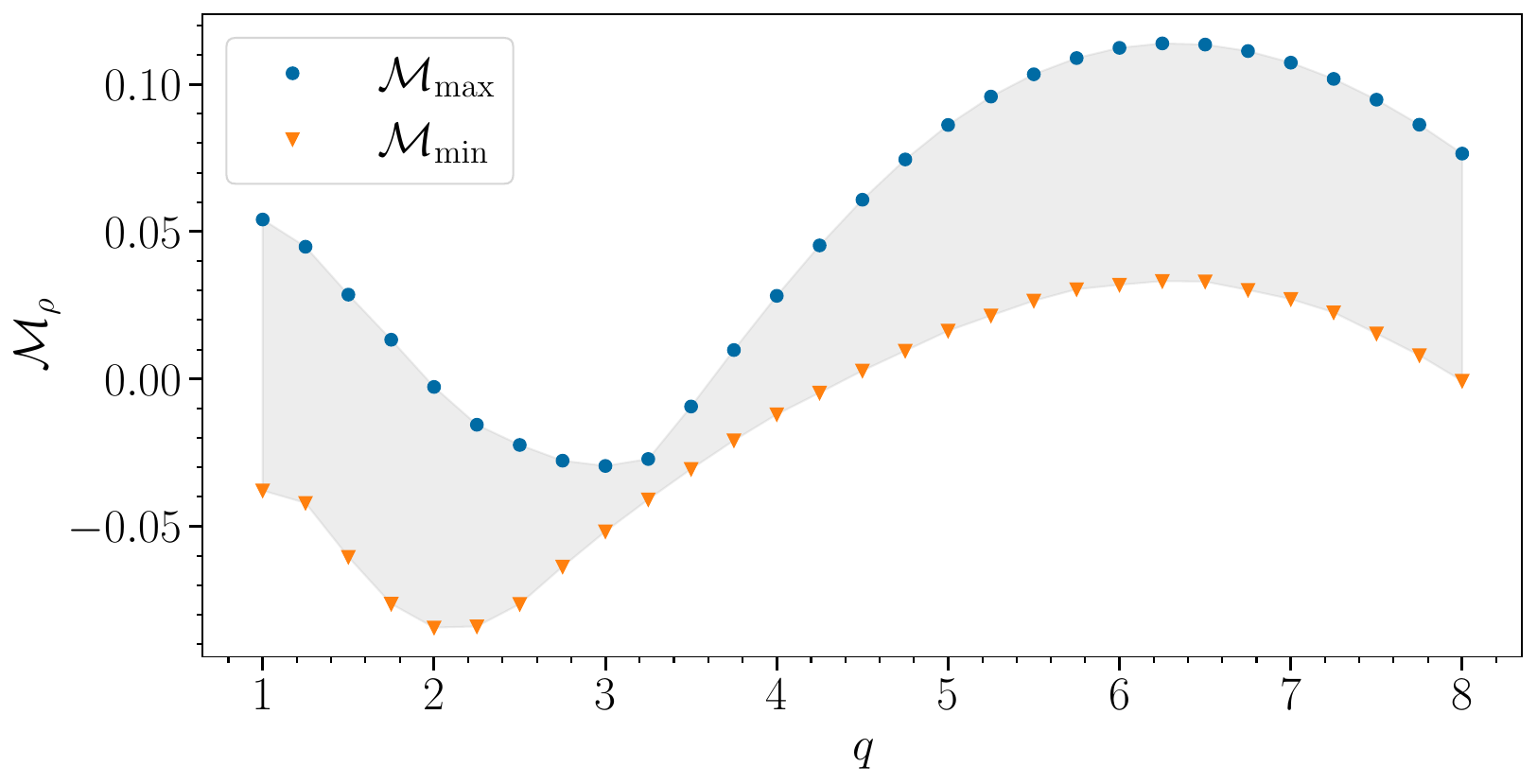}
    \caption{\textbf{SNR mismatch between the frequency-domain memory waveform model and the NR surrogate calculation}.
    As in Fig.~\ref{fig:mismatch}, the SNR mismatch was computed using the Advanced LIGO design-sensitivity noise curve.
    \emph{Top}: $\mathcal{M}_\rho$ in Eq.~\eqref{eq:amplitude_mismatch} as a function of primary mass $m_1 \in [5\,M_\odot,100\,M_\odot]$ for mass ratios $q = 1, 3, 6, 8$, with $m_2 = m_1/q$ constrained to the same range of masses as $m_1$.
    \emph{Bottom}: Maximum (blue circles) and minimum (orange triangles) SNR mismatch over $m_1 \in [5,100]M_\odot$ as a function of the mass ratio $q\in[1,8]$.
    The gray shaded region at fixed $q$ spans the range of values of $m_1$. 
    The SNR mismatch remains within $|\mathcal{M}_\rho| \lesssim 0.1$ across the full parameter space explored, which is comparable to the related results in Paper~II.
    } 
    \label{fig:amplitude_mismatch}
\end{figure}

The top panel of Fig.~\ref{fig:amplitude_mismatch} shows the SNR mismatch as a function of the primary mass $m_1$ for the same mass ratios displayed in Fig.~\ref{fig:mismatch}.
The bottom panel shows the range of SNR mismatch values at each $q$ different values of $m_1$ (analogously to the results in the bottom panel of Fig.~\ref{fig:mismatch} for the usual notion of the mismatch).
Unlike the mismatch in Fig.~\ref{fig:mismatch}, the SNR mismatch in Fig.~\ref{fig:amplitude_mismatch} is not a monotonic function of the primary mass (and by its definition, it can either be positive or negative).
Thus, depending on the values of $m_1$ and $q$, the phenomenological model can either underestimate or overestimate the optimal SNR of the surrogate memory signal by at most about ten percent.
For higher mass ratios, however, the SNR mismatch is primarily positive (meaning that it underestimates the SNR), and it tends to increase with the primary mass.
This is likely related to the fact that in Fig.~\ref{fig:memory_amp}, the amplitude model at mass ratio $q=8$ is consistently a small amount lower than the same amplitude computed from the surrogate model.
Compared with the model of Paper~II, the SNR mismatch is similar in magnitude.

\section{Conclusions} 
\label{sec:conclusions}

In this paper, we constructed a phenomenological frequency-domain model for the dominant $(l,m)=(2,0)$ mode of the gravitational-wave memory signal from nonspinning binary-black-hole mergers.
The earlier papers in this series on modeling the GW memory signal were focused more on the final memory offset and extreme mass-ratio limit (Paper~I~\cite{Elhashash:2024thm}) and a time-domain model and its analytical Fourier transform (Paper~II~\cite{Elhashash:2025hqi}).
Unlike the analytical Fourier transform of the time-domain model in Paper~II, the frequency-domain model presented here is more accurate and better optimized for frequency-domain calculations and analyses.

Although the time-domain memory signal is real-valued, its Fourier transform is complex-valued.
We constructed the full frequency-domain GW signal by modeling its complex amplitude and phase separately.
The amplitude model captures qualitative frequency-domain features of the memory signal, including the $1/f$ behavior at low frequencies associated with the nonzero final memory offset and its rapid decay at higher frequencies above the inverse rise time associated with the memory.
The phase was modeled independently using a phenomenological ansatz that captures the order-one changes in the residual phase that arise as a function of frequency and mass ratio.

The amplitude and phase models were calibrated for nonspinning BBHs with mass ratios from one to eight by using results computed using a hybridized NR surrogate.
There were free coefficients in the amplitude and phase models that were optimized in this calibration step through a nonlinear minimization procedure.
The combined amplitude and phase models give the full frequency-domain representation of the memory waveform that reproduced the surrogate signal this parameter space to good accuracy.

Specifically, we assessed the accuracy of the model by computing the mismatch between the phenomenological model and the GW memory signal computed using the surrogate model.
We used the Advanced LIGO design-sensitivity noise curve for this analysis.
The mismatch remains below approximately $4\times 10^{-3}$ across the full parameter space.
We also computed an SNR mismatch between the model and the surrogate, which is a better assessment of the quality of the amplitude model.
In this case, the SNR mismatch was at most roughly $10^{-1}$ across the parameter space.
Because the frequency-domain model of this paper also avoided the spectral artifacts associated with the finite differentiability of the time-domain model of Paper~II, the mismatch for the new model was improved by almost an order of magnitude from the mismatch for the model in Paper~II.
Moreover, the model developed here was significantly more efficient computationally than transforming the time-domain model in Paper~II to the frequency domain (on average around 30 times faster).

It would be of interest to generalize this model and method to cover a larger range of the parameter space of BBH binaries (specifically those including BH spins).
It would be most natural to first generalize the result to aligned-spin binaries before considering the larger parameter space of precessing BBH systems.

\acknowledgments

D.A.N.\ was supported in part by the NSF grant PHY-2309021 and the NSF-CAREER Award PHY-2439893.

\bibliography{refs}

\end{document}